\PassOptionsToPackage{bookmarksnumbered,unicode}{hyperref}
\documentclass[sigconf, nohyperxmp]{acmart}

\AtBeginDocument{%
  \providecommand\BibTeX{{%
    \normalfont B\kern-0.5em{\scshape i\kern-0.25em b}\kern-0.8em\TeX}}}

\setcopyright{none}
\usepackage[title]{appendix}%
\usepackage[graphicx]{realboxes}
\usepackage{xcolor}

\newcommand{\review}[1]{{\leavevmode\color{black}{#1}}}
\usepackage{caption}
\usepackage{subcaption}
\usepackage{placeins}
\usepackage{changepage} 
\usepackage{lscape}  
\usepackage{rotating} 
\usepackage{adjustbox}
\usepackage{longtable}
\usepackage{booktabs}

\usepackage{verbatim}

\begin{document}

\title[Barriers and Mitigation Strategies of SE with Non-traditional Backgrounds]{Towards Understanding Barriers and Mitigation Strategies of Software Engineers with Non-traditional Educational and Occupational Backgrounds}

\acmConference[Empirical Software Engineering 29 (4)]{Empirical Software Engineering 29 (4)}{1-47}{Author Version}
\acmYear{2024}
\acmDOI{10.1007/s10664-024-10493-1}
\setcopyright{none}
\acmISBN{}
\acmPrice{}
\author{Tavian Barnes}
\authornote{Equal contributors by alphabetical order}
\email{tbarnes@uwaterloo.ca}
\author{Ken Jen Lee}
\authornotemark[2]
\email{kenjen.lee@uwaterloo.ca}
\author{Cristina Tavares}
\authornotemark[2]
\email{cristina.tavares@uwaterloo.ca}
\affiliation{%
  \institution{University of Waterloo}
  \streetaddress{200 University Ave W}
  \city{Waterloo}
  \state{Ontario}
  \country{Canada}
  \postcode{N2L 3G1}
}

\author{Gema Rodríguez-Pérez}
\email{gema.rodriguezperez@ubc.ca}
\affiliation{%
  \institution{University of British Columbia}
  \city{Kelowna}
  \state{British Columbia}
  \country{Canada}}

\author{Meiyappan Nagappan}
\email{mei.nagappan@uwaterloo.ca}
\affiliation{%
  \institution{University of Waterloo}
  \streetaddress{200 University Ave W}
  \city{Waterloo}
  \state{Ontario}
  \country{Canada}
  \postcode{N2L 3G1}
}

\begin{abstract}
The traditional path to a software engineering career usually involves a post-secondary diploma in Software Engineering, Computer Science, or a related field. However, many individuals working as software engineers take a non-traditional path to their careers, starting from other industries or fields of study. This paper explores the barriers that individuals with non-traditional educational and occupational backgrounds face when pursuing a software engineering career and proposes potential strategies to overcome those barriers. A two-stage methodology was used, consisting of an exploratory study followed by a follow-up survey. The exploratory study consisted of a grounded-theory-based qualitative analysis of relevant Reddit data to yield a framework around the barriers and possible mitigation strategies. These findings were then supplemented through a follow-up survey. Understanding these barriers and what strategies could be effective is an important step towards making software engineering more accessible to individuals with non-traditional backgrounds. In addition to fostering functional diversity, this might also serve to tackle labor shortages within the software engineering industry.
\end{abstract}


\maketitle
\section{Introduction}\label{sec:intro}

The Software Engineering (SE) industry has seen significant annual growth in North America over the last decade~\cite{canstatsgrowth,ibisworld}. Despite the increasing number of Computer Science (CS) and SE degrees being awarded~\cite{datausacs, datausacse}, there is still a shortage of software engineers~\cite{hyrynsalmi2019motivates,forbesShortage}. This discrepancy in labor supply has drawn attention to alternative pathways to the traditional SE career, which conventionally involves obtaining a university degree in CS or a related field \cite{wilson2017building}. 
\review{Specifically, in this work, we define software engineers with a ~\emph{traditional background} as those who initiate their careers after completing a post-secondary program in SE or CS. On the other hand, a ~\emph{non-traditional background} encompasses software engineers who did not follow this traditional career path, including individuals who transitioned to SE from another profession, those who pursued a different field before SE, or those who acquired SE skills through self-learning.}

Success in the computing or SE field may not always require following the traditional SE route. Practical experience and strong coding skills can often hold more significance and value than solely possessing a degree. In fact, according to a survey conducted by So et al. \cite{so2020}, only 9.7\% of surveyed professionals working in SE-related positions consider a university degree as ``critically important'' for success in this domain. Furthermore, open-source projects developed by contributors with diverse backgrounds and varying levels of expertise are widely adopted and utilized by millions of people globally in modern society. Hence, to drive technological advancements, the importance of hands-on experience and coding skills may outweigh the significance of a traditional software engineering background.

In recent years, big tech companies such as Google, Apple, and IBM no longer require candidates to hold university or college degrees to be employed~\cite{cnbc18}. However, the viability and effectiveness of transitioning to a SE career for individuals with non-traditional educational and/or occupational backgrounds remain poorly understood. Academic diversity can be considered a type of functional diversity \cite{Harrison2007WhatsTD}, and although there is limited research on the effects of educational background diversity in the field of CS/SE, previous studies in other domains have revealed several advantages. For instance, having a diverse range of educational backgrounds can help prevent the ``groupthink'' mentality \cite{janis1972victims} and is also statistically correlated with increased innovation in businesses in eight different countries (USA, France, Germany, China, Brazil, India, Switzerland, and Austria) \cite{lorenzo2018and}.

This study aims to contribute a first step towards understanding the challenges encountered by individuals with non-traditional occupational and/or educational backgrounds when transitioning to a career in SE. Additionally, the research delves into potential strategies to alleviate these barriers. Gaining insights into these obstacles and mitigation approaches is crucial for fostering greater accessibility to SE careers for those with non-traditional backgrounds. This not only has the potential to address the existing labor shortage but also fosters a more diverse and inclusive SE community.

Hence, this paper poses the following research questions:
\begin{itemize}
    \item[\textbf{RQ1}] 
    \textbf{What are the perceived barriers faced by those with a non-traditional educational and occupational background who are either software engineers or attempting to switch into a SE career?}\\
    \textbf{Motivation:} 
    While a conventional SE educational background may not necessarily guarantee success in the SE industry as these roles frequently demand competencies beyond the focus of SE curricula (e.g., programming skills)\cite{karunasekera2007preparing}, there are numerous individuals educated or employed in other fields who possess a deep interest in SE and aspire to pursue a career in this domain. However, various obstacles may prevent them from doing so. Thus, to enhance diversity and inclusion in SE and to fully reap the benefits of having software engineers with diverse educational and professional backgrounds, it is essential to identify and comprehend these barriers.
    \item[\textbf{RQ2}] \textbf{What are the mitigation strategies that could lower reported barriers to success?}\\
    \textbf{Motivation:} Barriers are only half the story; a comprehensive understanding also includes exploring the mitigation strategies currently employed by those facing these barriers. Exploring RQ2 can provide insights into the effectiveness of each barrier's mitigation, the roles different parties can play in facilitating existing mitigation strategies, and the potential to empower individuals with non-traditional backgrounds with novel mitigation tactics that are not currently employed.

\end{itemize}

\review{We used a mixed-method approach to answer our RQs. We first performed an exploratory study, using a grounded-theory-based approach to analyze more than $133,000$ relevant posts from Reddit. With the exploratory study, we aim to identify the barriers and mitigation strategies reported by Reddit users with non-traditional education and occupational backgrounds. This analysis resulted in more than 30 barriers and strategies. We then complement the qualitative study with a survey of 46 participants to better understand how relevant each barrier and strategy is, and how their relevance compare to each other. Moreover, survey participants also reported various unique strengths of having non-traditional backgrounds. Finally, this paper proposes a theoretical model that is built by combining the results from the qualitative and quantitative studies. This model separates the transition process into four phases, and presents which types of barriers and strategies are relevant for each phase, and how non-traditional backgrounds can be beneficial.}

The rest of the paper is organized as follows. Section~\ref{sec:relevant_work} discusses the related work. Section~\ref{sec:meth} presents our exploratory and follow-up survey design, including data collection, data analysis, and theory development. Section~\ref{sec:find} answers our research questions and shows the findings of our study. Following that, Section \ref{sec:model} presents a theoretical model, while Section~\ref{sec:disc} discusses the results. Finally, Section~\ref{sec:thre} highlights the threats to validity, and Section~\ref{sec:conc} concludes the paper and discusses future work.

\section{Relevant Work}
\label{sec:relevant_work}

In this section, we review relevant research on diversity in SE, diversity in terms of educational and occupational backgrounds, and the career switching process.

\subsection{Understanding Diversity in CS/SE}
Harrison and Klein \cite{Harrison2007WhatsTD} define diversity as the ``distribution of differences among the members of a unit concerning a common attribute.'' Most of the previous diversity research in SE focuses on perceptible attributes like gender, age, and ethnicity, especially as these are protected classes under the United States anti-discrimination law \cite{EeocProtected}. Previous works have shown that software practitioners face challenges and barriers to developing activities in this field resulting from perceived diversity related to these attributes \cite{Rodriguez-PerezGema2021Pdis}.
Age-related barriers are particularly relevant to software engineers with non-traditional academic backgrounds, who tend to be older at the same career stage as their peers \cite{CourseReport}. The perception that younger software engineers are better is pervasive throughout the industry, and even popular media \cite{paper5}. This discriminatory discourse tends to create challenges for older software engineers to participate in Open-Source Software (OSS) projects \cite{morrison2016veteran,baltes202040}.

\subsection{Diversity in Educational and Occupational Backgrounds}
In team settings, educational background diversity refers to the different sets of knowledge, skills, and abilities that team members acquire. Thus, a team composed of people with diverse educational backgrounds can understand a problem based on various combinations of information, insights,
and perspectives~\cite{LuanKun2016Tneo}. Studies conducted outside the SE domain have revealed that diverse educational backgrounds may affect work outcomes and team behaviour in several aspects, such as team creativity, team performance, and innovation 
\cite{JehnKarenA1999WDMa,PhillipsKatherineW2004Dgai,GuoWeixiao2021WDEL, MohammadiAli2017WCaI}. 
However, an overly high educational background diversity can lead to an excessive conflict of ideas that is time-consuming due to the need for reconciling knowledge, experiences, and different views~\cite{LuanKun2016Tneo}. According to Kearney and Gebert, the formation of a team with diverse functional and educational profiles is a way to promote the integration of crossed ideas and present new perspectives and insights to the team to solve a problem or execute a task~\cite{KearneyEric2009MDaE}. Guo et al. demonstrate that educational diversity may negatively affect team creativity, especially when tasks are repetitive, defined, and predictable, and the frequency of team personnel changes is higher \cite{GuoWeixiao2021WDEL, Chen2018TheRO}. Cases of research have also shown that educational diversity influences, at different levels, team performance in the banking and financial sectors 
\cite{rizwan2016impact}. 

Functional background diversity refers to differences in the professional background among team members \cite{RichterAndreasW2012CSaI,BundersonJ.S2002CACo,Chen2018TheRO}.
Previous findings indicate that functional diversity is associated with the promotion of creativity, innovation, performance, and problem-solving ability. It also allows a high capacity for incorporating new knowledge from a diverse domain \cite{SimonsStarleneM2011Daii, SomechAnit2006TEoL}.
Although diverse functional backgrounds offer teams various benefits, different views and perspectives might cause disagreements among team members, negatively impacting teams \cite{SomechAnit2006TEoL}.
Also, functional background diversity may be negatively impacted by rising costs associated with the time-consuming consensus process when integrating ideas~\cite{SimonsStarleneM2011Daii}.

\subsection{Career Switching}\label{sec:careerswitching}

While prior research on career transitions related to the software industry is limited, Castro et al.'s study on career transitions by junior academics into Data Science is of particular relevance \cite{RuizCastro2020}. Junior academics face career barriers within academics, including career uncertainty and the lack of meaning and impact. These barriers, when paired with a failure to meet career expectations, motivated a career switch. The transitioning process is facilitated by what Castro et al.~\cite{RuizCastro2020} referred to as \textit{career catalysts}, which could take the form of career training workshops, development centers, bootcamps, etc. These career catalysts could yield three key benefits, including adapting one's mindset to new industries and roles, understanding potential future jobs and employers, and gaining meaningful employment. 
Career sustainability, in turn, was experienced in two ways, i) a reaffirmation of professional identity through the nurturing of characteristics relevant to the academic researcher role while working as a data scientist, and ii) the reconciliation of the professional identity, through the fulfillment of positive expectations (that were previously unmet in the researcher role) while working as a data scientist \cite{RuizCastro2020}.

Our work goes one step further and does not focus on junior academics willing to transition to Data Science. It aims to contribute to the current literature with a more high-level understanding of the barriers faced by, and mitigation strategies used by, those with any non-traditional occupational and/or educational background transitioning into any role related to SE. To the best of our knowledge, our paper is the first work that studies the reported barriers and potential mitigation strategies to make SE jobs more accessible to those with non-traditional backgrounds.

\review{Following STGT's recommendation of a targeted literature review stage, (discussed later in Section \ref{sec:stgtadvancedstage}), we investigated previous works on the career transition process when it slowly became clear during data analysis that barriers and mitigation strategies for transitioning into SE is a process-based phenomenon. A particularly relevant existing model is the Integrated Career Change Model by Rhodes and Doering \cite{Rhodesmodel}. Derived from many theories that have been empirically supported, the model is a flow diagram consisting of 17 components that provide a framework of one's motivation behihe career changes, and the process of the change itself. The model has since been widely cited, having more than 300 citations based on Google Scholar as of the time of writing. Using this model, Carless and Arnup \cite{Carless_2011} found several individual characteristics to be associated with career change. Higgins \cite{Higgins_2001} showed that beyond individual characteristics, an individual's choices about career changes are socially embedded. While there exist other career change models, e.g., Hind \cite{hind2005making} used Janssen's ``The Four Roomed Apartment of Change''\cite{janssen2005introduction} to explore career changes, Rhodes and Doering's paper \cite{Rhodesmodel} is the most suitable for our context. }

\section{Methodology}
\label{sec:meth}

To answer our research questions, we performed an exploratory study where we qualitatively analyzed Reddit posts by our target population: those with non-traditional educational or occupational backgrounds, i) who are considering a SE career, or ii) who have already started their SE career. 
During the exploratory analysis, we followed Hoda's Socio-Technical Grounded Theory (STGT) \cite{Hoda2021} to uncover the possible barriers and mitigation strategies. Then, to better understand the relevance of the barriers and mitigation strategies gathered from the qualitative analysis, we surveyed our target population.

\begin{figure}[ht]
  \centering
  \includegraphics[width=\linewidth]{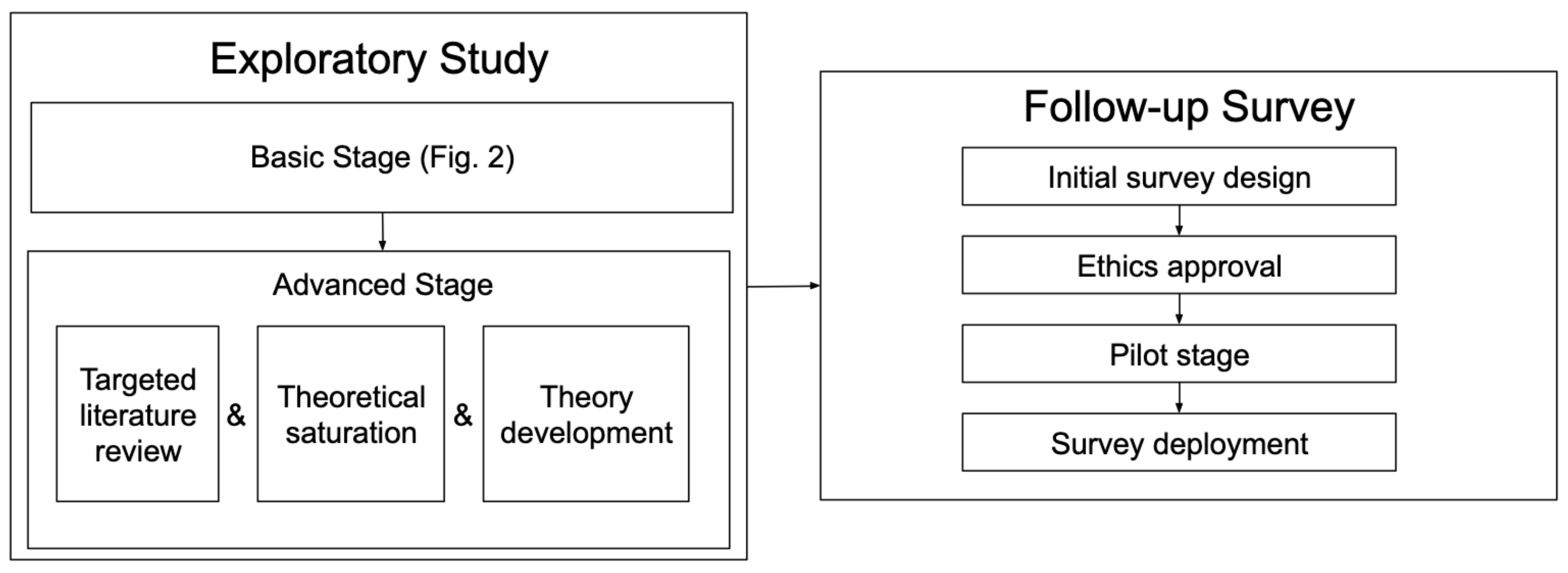}
  \caption{A high-level overview of the methodology of this work, which contains an exploratory study followed by a follow-up survey.}
  \label{fig:methodology}
\end{figure}

A high-level overview of our methodology is presented in Fig. \ref{fig:methodology}. The exploratory study consists of a STGT-based basic stage and a STGT-based advanced stage. Below, we discuss the exploratory study, its stages and the follow-up survey in detail. \review{A previous version of the methodology was submitted as part of a Registered Report \cite{barnes2022understanding}.}

\subsection{Exploratory Study}\label{sec:method:exploratory}

Our exploratory study uses a qualitative analysis process inspired by the STGT, which consists of two stages: i) the \textit{Basic Stage} and ii) the \textit{Advanced Stage} (Fig. \ref{fig:methodology}). Hoda's STGT~\cite{Hoda2021} is ``an iterative and incremental research method for conducting socio-technical research using traditional and modern research techniques to generate novel, useful, parsimonious, and modifiable theories.'' Hoda's STGT aligns very well with our study for two reasons. First is the socio-technical nature of our topic, i.e., the barriers faced by those who want to transfer into a SE career. Second, STGT is designed with a wide variety of research methodologies in mind, which allow for context-specific ontological stands, including non-physical realities like Reddit, a virtual world. Many existing versions of Grounded Theory (GT) are ``predominantly applied as a qualitative method through traditional data collection techniques such as interviews and observation'' \cite{Hoda2021}.

However, some aspects of our exploratory study's procedures deviate from the traditional STGT.
While STGT takes inspiration from Glaserian, Strauss-Corbinian, and Constructivist versions of GT, Hoda \cite{Hoda2021} does not specify: i) the necessity of defining research questions before starting data collection and analysis, and ii) the potential utilization of Inter-Coder Reliability (ICR) statistics to assess the reliability of analysis conducted by multiple coders.
Our study defined research questions prior to the analysis, closer to Strauss-Corbinian GT, in which ``a question should be pre-set as it sets the boundaries around the study area'' \cite{Coleman2007}.
Moreover, the RQs are structured in terms of barriers and mitigation strategies, which is a structure commonly used in existing SE research on various aspects of diversity (e.g., \cite{ford2016paradise,canedo2021breaking,paper5,kaur2014agile,sharma2015communication}). This way of structuring RQs was done to strike ``a balance between being sufficiently informed versus overly influenced by existing works'', a Constructivist GT approach \cite{Hoda2021}.
Regarding ICR statistics, we describe below the procedure we used for incorporating Krippendorff's Alpha \cite{krippendorff2011computing} to ensure reliability.
Finally, this work had pre-determined data sources (i.e., Reddit) instead of allowing for the possibility of theoretical sampling from other unplanned data sources \cite{Hoda2021}.  We believe that Reddit serves as a rich data source sufficient for answering our research questions.

\textbf{Ethics Considerations.} Since this study involves the analysis of a publicly available archive of Reddit
posts hosted by pushshift.io \cite{pushshift}, ethical considerations have been applied. On this, Gold and Krinke demonstrate the use of Menlo ethics principles within Mining Software Repository (MSR) research contexts~\cite{Gold2021}. Since obtaining informed consent from Reddit users whose content is analysed in this work is logistically impossible, we will not present any direct quotations from the data. We will anonymize the data by removing usernames and make it privately available for any future researchers interested in replicating or building off of this work. This aligns with Hoda's suggestion to ``obscure identifiable information'' when using public data \cite{Hoda2021}.

\subsubsection{Basic Stage}

\begin{figure}[!ht]
  \centering
  \includegraphics[width=.7\linewidth]{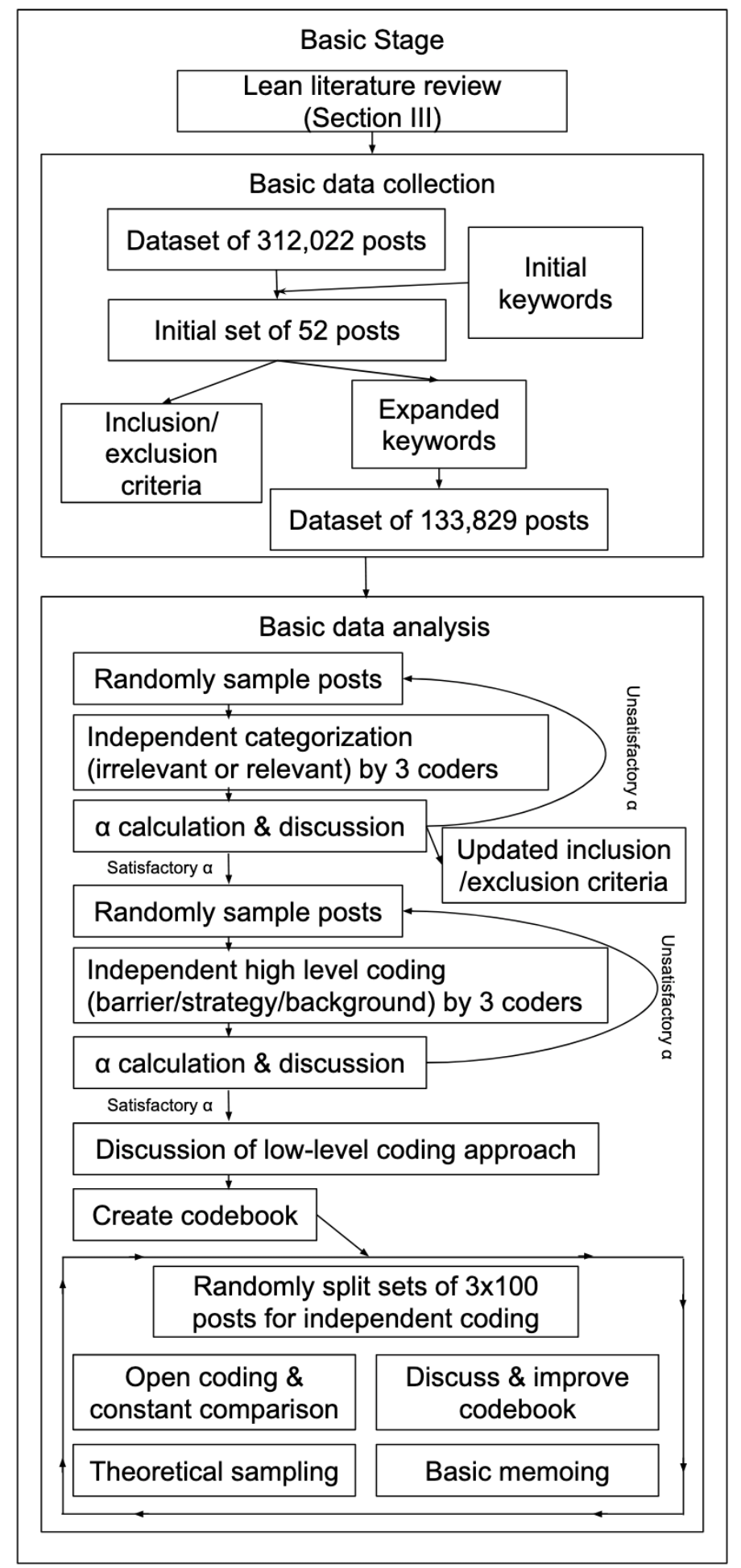}
  \caption{An overview of the Basic Stage which encompasses: lean literature review, basic data collection, and basic data analysis.}
  \label{fig:methodologyBasic}
\end{figure}

This stage involves a lean literature review, basic data collection, and basic data analysis \cite{Hoda2021} (Fig.~\ref{fig:methodologyBasic}).

\textbf{Lean Literature Review.} The purpose of performing a lean literature review is to ``identify gaps and motivate the need for a study'' \cite{Hoda2021}. Findings from our lean literature review have been presented in Section \ref{sec:relevant_work}. These findings revealed that there is no previous work that explores our research questions. This motivates our study, especially given the significance of our research questions, as explained in Section \ref{sec:intro}. To borrow Hoda's words, the topics explored in this paper are ``relatively nascent, with no or few existing theories'' \cite{Hoda2021}.

\textbf{Basic Data Collection.} 
We chose Reddit as our data source because: i) it is widely used and one of the top 20 most popular websites in the world \cite{amaya2021new}; and ii) it has subreddits, which are forums on Reddit for discussions on specific topics, with posts that are relevant to our research questions. 
Our dataset was collected from Reddit posts made in the three-year period from 2017 to 2019, using a publicly available archive of Reddit posts hosted by pushshift.io \cite{pushshift}. We wanted to focus on data prior to 2020 to avoid any confounding variables introduced by the global pandemic.
On the other hand, to select an appropriate start year, due to the lack of a clear way of studying trends in the SE world, we observed trends in the use of programming languages as reported by Stack Overflow's annual survey. Particularly, 2017 was when the use of Python increased significantly from 25.9\% in 2016 \cite{so2016} to 32\% in 2017 \cite{so2017} among all survey responses. As such, we chose 2017 as the start year.
We restricted the data to a set of seven subreddits, denoted by ``/r/", where posts relevant to our study would be on-topic:~\texttt{ /r/\{\textit{learnprogramming, AskProgramming, cscareerquestions, SoftwareEngineering, cscareerquestionsEU, ExperiencedDevs, and codingbootcamp}\}}. The selection of these specific subreddits was based on their relevance, considering they were among the most significant ones out of the top $1,000$ subreddits with the highest number of subscribers \cite{subredditstats}. As a result, the dataset comprises a total of $312,022$ posts.

To find potentially relevant posts within this dataset, we built a simple keyword search engine.~\footnote{The source code of the search engine is available at \url{https://github.com/tavianator/pheddit}} The search engine is written in 
Rust and performs a full-text keyword search over the titles and post bodies of every Reddit post in the dataset.
The search is case-insensitive and respects word boundaries, but does not perform any more advanced query processing such as stemming. We compensated for this by explicitly including relevant pluralization and conjugations of our keywords.

Using this search engine, we iteratively refined a set of search queries that gave a large number of relevant documents. Starting from an initial set of queries, the first three authors worked through the resulting posts to determine whether they deemed them relevant, referring to and revising their inclusion and exclusion criteria as necessary. The authors also analyzed whether the post contained keywords that might be useful for future search queries. 

\begin{figure*}
\centering
\begin{minipage}{.8\textwidth}
  \centering
  \includegraphics[width=.9\linewidth]{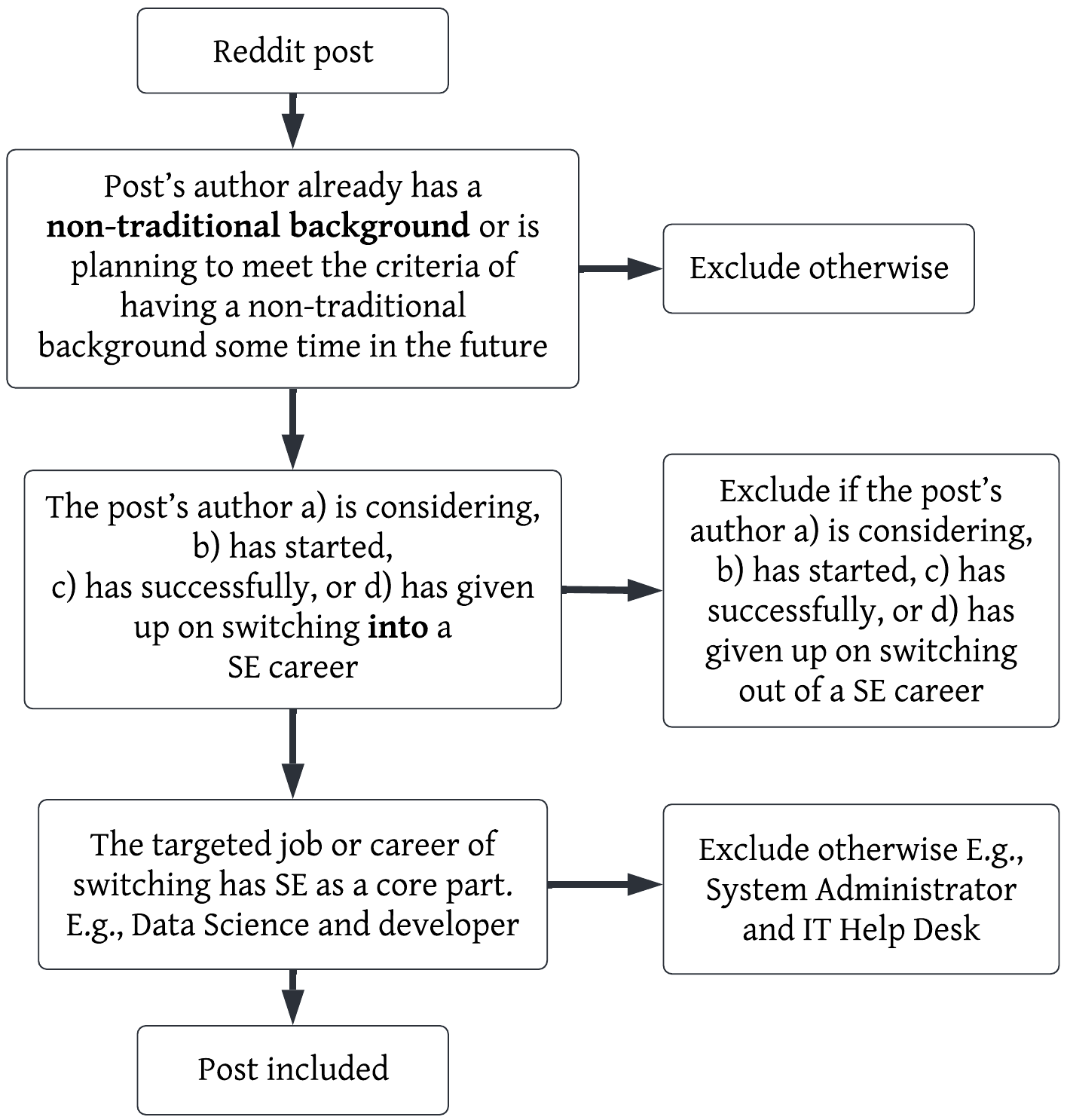}
  \captionof{figure}{Initial Inclusion/Exclusion Criteria}
  \label{fig:incl_exc_init}
\end{minipage}%
\end{figure*}

A Reddit post is included for qualitative analysis (i.e., considered relevant) if it meets the inclusion and exclusion criteria. To generate these criteria, the first three authors individually gathered a total set of 52 posts using an initial set of keywords that they thought were relevant to the research questions. These keywords included \textit{career}, \textit{career change}, \textit{self-taught}, \textit{switch software engineering}, \textit{back to school}, and \textit{community college}. The posts were also chosen to aid in the process of designing the inclusion and exclusion criteria, i.e., there were posts obviously relevant or irrelevant, and posts that were ambiguous. Then, as a group, we generated the inclusion and exclusion criteria by discussing the relevance of each post (Fig. \ref{fig:incl_exc_init}); which were intended to be used to judge a post's relevance based only on the information available in the post itself (i.e., not including any other information like profile information or other posts by a post's author).
The first criterion checks if authors of included posts either have non-traditional backgrounds already, or plan to become someone with a non-traditional background in the future. An example that satisfies the latter condition is if someone who is currently completing a CS undergraduate degree creates a post asking about how easy it is to pursue a SE carer if they dropped out of university (hence becoming someone with a non-traditional background due to having no relevant graduate degrees).
The second criterion enforces directionality; a post author should be on the path of transitioning \textbf{into}, instead of out of, a SE career.
Lastly, the third criterion ensures that the type of jobs that a post's author is thinking of transitioning into involves SE as a core part of the jobs' tasks.

In general, our motivation is to cast a reasonably wide net when gathering the various barriers and mitigation strategies; i.e., we do not want to have a narrow scope that artificially excludes the messiness and complexity of the phenomenon studied. As such, other trickier cases (e.g., someone who started their university education in CS and switched to an irrelevant area halfway through) will be accessed on a case-by-case basis based on their contribution towards our process of answering the research questions.

Based on the initial set of posts, the first three authors performed a keyword expansion exercise as a group to generate a final set of keywords that reflects relevant concepts to our research questions.
The final set of keywords contains 12 words, including \textit{degree}, \textit{career}, \textit{programming}, \textit{school}, \textit{learn}, \textit{switch}, \textit{change}, \textit{college}, \textit{university}, \textit{advice}, \textit{bootcamp}, \textit{self-taught}. These keywords were used to search the title and body content (i.e., excluding the comments and replies) of all posts in our data pool, resulting in $133,829$ possibly relevant posts. 

\begin{figure*}
\begin{minipage}{.8\textwidth}
  \centering
  \includegraphics[width=.9\linewidth]{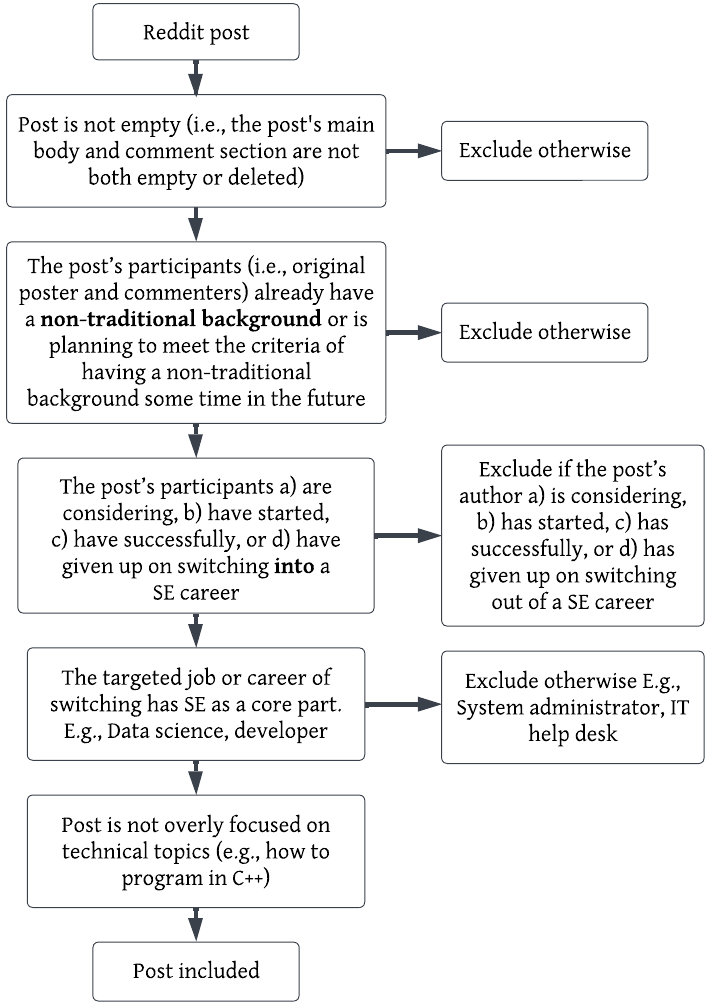}
  \captionof{figure}{Updated Inclusion/Exclusion Criteria}
  \label{fig:incl_exc_final}
\end{minipage}
\end{figure*}

\textbf{Basic Data Analysis.} This step involves three main steps, namely post-categorization (i.e., is a post relevant or not), deductive coding (i.e., identifying barriers, strategies, and backgrounds), and open coding (i.e., for specific types of barriers, strategies, and backgrounds). In this study, we used NVivo 12, a data analysis software commonly used in qualitative contexts (e.g., \cite{tkalich2021making,Masood2020}), to aid with various steps in our qualitative analyses, including open coding and memoing.
During qualitative coding, coders (i.e., the first three authors) treated each comment or the original post as a single unit of coding, a common practice for Reddit-based qualitative coding (e.g., \cite{Valle2021}). 
To ensure an acceptable level of reliability between the three coders, their intercoder-reliability (ICR) was measured at the first two stages of coding: i) when categorizing posts as relevant or irrelevant based on the inclusion and exclusion criteria, and ii) when deductively coding for the high-level codes of barriers, strategies and background.
Following that, inductive open coding was used to identify the specific types of barriers, strategies, and background; ICR was not used at this stage, as shown in Fig. \ref{fig:methodologyBasic}.

As illustrated in Fig. \ref{fig:methodology}, the first step was to ensure all coders had a good understanding of how to categorize posts as relevant or irrelevant. 
Posts were randomly sampled and independently categorized by all three coders. Krippendorff's Alpha \cite{krippendorff2011computing} was then calculated to measure ICR. 
Calculating ICR is important for this step of categorizing each post's relevance to ensure a reliable level of common understanding of the inclusion/exclusion criteria among the coders, i.e., if a round of coding yields an unsatisfactory ICR, that is a sign that the criteria could be further clarified or improved.
Regarding the ICR metric, we chose Krippendorff's Alpha \cite{krippendorff2011computing} given its flexibility in regard to the number of coders, sample size, etc. \cite{hayes2007answering}, its suitability for our study's design \cite{Feng2013} and its use in existing works that qualitatively analyzed mined Reddit data \cite{kumar2018learning,sparrow2020silly}. 
In the first iteration, 100 posts were randomly sampled, and the coders achieved $\alpha = 40.9\%$. Then, 50 posts were randomly sampled and independently categorized in the second iteration, $\alpha = 66.9\%$, and the third set of 50 posts yielded a satisfactory $\alpha = 79.2\%$ (i.e., close enough to the $80\%$ threshold \cite{Lombard2002}).
The second and third iterations involved only 50 posts as the coders learned, after the first iteration, that having more frequent discussions based on smaller batches of posts was more productive towards establishing ICR.
Discussions during these iterations yielded a few refinements to the inclusion/exclusion criteria, as shown in Fig.~\ref{fig:incl_exc_final}. 
Particularly, the coders realized that there were a few posts with seemingly relevant titles, but are empty, often because the post's content has been deleted. As such, an extra criterion has been added to explicitly filter out empty posts.
The coders observed that some posts had relevant titles but were empty due to deleted content. To address this, a criterion was added to filter out empty posts. Additionally, relevance is now determined not just by the initial post but also by relevant content in the comments. The final criterion considers both the post's author and commenters.
The coders also discovered that posts from the studied population could be irrelevant if they only focus on technical topics. For instance, a post from someone transitioning into a software engineering career might ask about the best programming language or textbook for learning C++. In such cases, the non-traditional background of the post's author does not impact the technical discussions, and these discussions do not contribute new findings to the research questions.

After that, the next step was to have a reliable method of categorizing relevant content within each post as either barriers, mitigation strategies or background. Although this work is mainly interested in the barriers and strategies, a separate code was created for background (e.g., native language, age) since it often provides the context and lived experiences of a person's barriers and strategies. 
As such, this stage involves deductively coding using three previously determined high-level codes: barriers, strategies, and background.
The three coders coded two relevant posts each iteration and achieved satisfactory ICR ($\alpha \geq 80\%$ \cite{Lombard2002}) after four iterations (Table \ref{tab:icr}). Specifically, for each unit of coding, a binary number was assigned for each coder and each code. For example, if coder A coded a unit as containing only barriers, a 1 is assigned for barrier, and 0 for both strategy and background. These binaries were then used to calculate the Alpha value using R.
Although the identification and classification of barriers, strategies and background might seem too straightforward to justify the need for initial rounds of ICR calculations, the coders identified and subsequently resolved several differences in the understanding of these three categories through this process.
The main differences are related to ensuring any coded category (barrier/strategy/background) pertains to the work's targeted population and is relevant to the process of switching into a SE career. This was confusing at times because it is not always clear whether commenters in a post have non-traditional backgrounds themselves, or if their experience is specifically about transitioning into a SE career. 
The use of each comment as a separate unit of coding helped the coders with identifying these differences.

\begin{table}[htbp!]
\caption{Krippendorff's Alpha for each iteration of coding two posts' content for barriers, mitigation strategies, or background information.}
\begingroup
\begin{center}
\label{tab:icr}
\begin{tabular}{ p{0.1\linewidth} c c c }
   Iter. \# & Barrier & Mitigation Strategy & Background \\
  \hline
  1 & 54.4\% & 51.4\% & 34.5\%\\
  2$^*$ & 28.5\% & 3.8\% & 100\%\\
  3 & 51\% & 66.4\% & 70.6\%\\
  4 & 100\% & 83.2\% & 100\%\\
  \hline
\end{tabular}
\end{center}
\textit{* The agreement for this iteration was particularly low because two posts that were especially challenging to code were chosen. Specifically, these posts revealed confusion surrounding what barriers, strategies and backgrounds were of relevance to our research questions. For example, barriers should not be coded as barriers if they are due to the nature of a sector instead of transitioning into SE jobs in general.}
\endgroup
\end{table}

Following that, open coding for the specific types of barriers, strategies, and backgrounds was performed.
Three randomly chosen sets of 100 posts were created, and each set was individually coded by a coder. At the end of the coding process and analysis, it was determined that more data was needed, and a second set of 100 posts was created for each coder for further analysis. During the analysis, several methods outlined in STGT were employed. First was a constant comparison, which is ``the process of constantly comparing derived codes within the same source and across sources to identify key patterns in the data'' \cite{Hoda2021}.
We also performed basic memoing, a process where researchers document their thoughts, ideas, and reflections on emerging concepts and their connections during qualitative analysis \cite{Hoda2021}. For example, a relevant post described an individual who has been working temporary jobs feeling self-doubt about their abilities to switch into SE, since they perceive themselves as less capable than peers who have been working full-time jobs. A memo was created to note how this post, similar to a few other posts, touches upon the effects of an individual's social context on their own self-perception and intention to switch into SE.
Next, theoretical sampling, an ongoing sampling process that involves the intentional selection of data sources based on the data's specific characteristics \cite{Hoda2021}, was also employed. During the analysis, the coders continuously adapted their coding by paying more attention to data that contributed fresh perspectives.
Since this stage of inductive open coding was more dynamic, ICR was not calculated; instead, the coders aimed to achieve knowledge sharing and common understanding through group discussions.

We transitioned from the Basic Stage to the Advanced Stage upon the emergence of a few strong categories (e.g., barriers and strategies related to searching for SE jobs), as recommended by Hoda \cite{Hoda2021}. 

\subsubsection{Advanced Stage}\label{sec:stgtadvancedstage}

The Advanced Stage involves targeted literature review, theoretical saturation, and theory development (Fig.~\ref{fig:methodology}).

\textbf{Targeted Literature Review.} 
This refers to ``an in-depth review of literature targeting relevance to the emerging/emergent categories and hypotheses'' \cite{Hoda2021}. 
In our work, targeted literature reviews were done on existing career changing models (Section \ref{sec:careerswitching}).

\textbf{Theoretical Saturation.}
This refers to when new data collected no longer generate or contribute significantly to the existing findings (e.g., concepts, categories) \cite{Hoda2021}.
Practically, we defined our analysis as having reached theoretical saturation when during independent coding, each of the coders arrived at a point of analyzing three theoretically sampled Reddit posts that no longer contributed significantly to existing findings. This occurred during the analysis of each coder's second set of 100 posts (Fig. \ref{fig:methodologyBasic}).
The follow-up survey was used as a form of triangulation to facilitate an understanding of the trustworthiness of the STGT-based findings \cite{Aldiabat2018}.

\textbf{Theory Development.}
STGT provides researchers with a choice of two theory development approaches, \textit{Emergent Mode} or \textit{Structured Mode} \cite{Hoda2021}.
The Emergent mode is preferable when the basic stage reveals categories with emerging relationships, but without a clear theoretical structure. It involves targeted data collection and analysis (where the focus is paid almost exclusively to the most significant categories), and theoretical structuring (where researchers could explore how the emerging theory fits with theory genres and templates). On the other hand, the Structured mode is recommended for use when the Basic Stage yields categories with a relatively clear theoretical structure. It involves structured data collection and analysis (where relationships between key categories are strengthened), and theoretical integration (where categories are fully integrated into the overall structure of the framework). For our analysis, we employed the Structured mode due to the clear theoretical structures of the barriers, strategies, and backgrounds at a fundamental level. However, the integration of these categories required further clarification. Therefore, subsequent stages of analysis were specifically dedicated to exploring the deeper connections between them.

\subsection{Follow-up Survey}

The follow-up survey involves the design and administration of a survey that aims to supplement findings on barriers and mitigation strategies from the exploratory study (Fig.~\ref{fig:methodology}).

\textbf{Targeted Audience.}
To be eligible to take the survey, individuals must meet the following criteria: i) be at least 18 years old, ii) possess significant expertise (either a post-secondary degree/diploma or a minimum of three years of work experience) in a domain other than SE, and iii) currently employed as a software engineer, or currently studying SE, or have prior experience working in or studying SE subsequent to their involvement in another field.

\textbf{Survey Design.}
The survey asked participants about their demographics, educational background, and occupational background. The core part of the survey focused on validating findings from the exploratory study, which included barriers and mitigation strategies. 
Specifically, participants were asked to do the below in order:
\begin{itemize}
    \item Rate how relevant each barrier from the exploratory study is in their experience using Likert scales ($1 = $\textit{Not relevant at all} and $5 = $\textit{Extremely relevant})
    \item List any other barriers they faced 
    \item Rate how effective each mitigation strategy from the exploratory study is using Likert scales ($1 = $\textit{Not effective at all} and $5 = $\textit{Extremely effective})
    \item List any other mitigation strategies they used 
\end{itemize}

\textbf{Participant Recruitment.}
Participant recruitment was done via two main methods. First, emails were sent to two mailing lists within the authors' institution targeted at graduate students in areas related to Computer Science. We believe this to be an appropriate method, since our findings from the exploratory study (discussed below) revealed that going back to CS/SE graduate programs is one way of transitioning into a SE career. Moreover, it is not uncommon to come across graduate students that are part of those mailing lists who have non-traditional backgrounds, e.g., through taking classes, cooperating on other research projects, attending talks. The second method is through snowball sampling with personal connections that might fit the participation criteria.
The plan was to start recruiting participants through these two methods, before moving on to other methods, e.g., recruiting on relevant subreddits, if insufficient participants were recruited via the two main methods.
However, the survey received sufficient participants using the two main recruitment methods, so other methods were not employed in the end.
As such, we expect survey participants to be similar to Reddit users (the US accounted for 48.93\% of all Reddit traffic in the six months leading up to June 2021~\cite{reddittraffic}) analyzed in the Exploratory Study in that they are located in North America; i.e., participants recruited through the mailing lists are currently studying in a North American postsecondary institution, and those recruited through snowball sampling via personal connections are obtained by reaching out to those currently located in North America.
Each participant received a remuneration of CAD\$ 10 for completing the survey.

\textbf{Pilot study.} 
Following STGT's recommendations, we conducted a pilot study by running three participants through the survey using a think-out-loud methodology (see \cite{gupta2015identifying,latoza2006maintaining,10.1145/3511430.3511439} for similar survey pilot methodologies), and two researchers were present for each pilot. Feedback from these participants was used to improve the survey wording and structure. 
Pilot data were included in the analyses below because there were no significant changes in the survey after the pilot.
That said, there were still a few minor changes. For example, the initial survey design asked participants to list possible mitigation strategies on their own before asking them to rate mitigation strategies found in the exploratory study. The intention was to understand their perspective better without biasing them with the strategies found in the exploratory study.
However, during the pilot, it was obvious that participants were more fatigued due to having written many strategies that overlap with strategies in the given list. As such, the order was swapped in the final survey, which has been made available on Zenodo.~\footnote{Please refer to \url{https://zenodo.org/records/10511167} for the final survey used in this study.} The pilot study responses were received between January 31, 2023, and February 3, 2023, with the rest of the responses received between February 15, 2023, and March 8, 2023.

\textbf{Ethics Considerations.} Approval from our institution's ethics committee was obtained before the pilot and actual survey deployment. Moreover, we did not force a response for any of the questions. We also separated the collection of the anonymous response from the personally identifiable information needed for remuneration through two different Qualtrics forms. When participants who declared that they meet the selection criteria provided consent and completed the survey, they were automatically redirected to the second form that collects information needed for remuneration. 

\section{Findings}
\label{sec:find}

In this section, we present findings from both the exploratory study and the follow-up survey. 
\subsection{Exploratory Study}\label{sec:resultsexplore}

\begin{table*}
\caption{List of barriers and mitigation strategies. Applicable stages (Fig. \ref{fig:model}) for each are placed in brackets: \textit{I Intention to change career; II Preparation to change; III Job search; IV Job attained}. Additionally, strategies that are especially relevant for certain barriers are placed in brackets as well for their respective barriers.}
\label{tab:barriersNstrats}
\begin{tabular}{p{.18cm}p{7.9cm}p{.2cm}p{8.2cm}}
\toprule
ID & Barrier  & ID & Mitigation Strategy\\
\midrule
\multicolumn{4}{c}{\textbf{Education}} \\
B1 & Restricted access to desired educational resources [II; S1-6] & S1 & Attending university [II]  \\
B2 & Poor quality of educational resources [II; S1-6]  & S1.1 & In-person undergraduate programs \\
B3 & Difficulties with comprehending learning materials [II]  & S1.2 & In-person graduate programs \\
B4 & Uncertainty planning an educational pathway [II; S1-8] & S1.3 & Online undergraduate programs \\ 
B5 & Uncertainty with choosing educational resources [II; S7] & S1.4 & Online graduate programs \\
&& S2 & Attending a bootcamp [II] \\
&& S3 & Attending community college [II]\\
&& S4 & Self-learning [II] \\
&& S4.1 & Begin by learning web development\\
&& S4.2 & Avoid being overwhelmed by details when self-learning\\
&& S5 & Finding tutoring [II]\\
&& S6 & Getting professional certificates [II]\\
&& S7 & Choosing educational resources with better personal fit [II] \\
&& S8 & Choosing a reputable education path [II] \\
\midrule
\multicolumn{4}{c}{\textbf{Financial}} \\
B6 & Financial cost of software engineering education [II; S9] & S9 & Getting financial aid [II] \\
B7 & Financial cost related to resources for job search [III; S9]  &  & \\
\midrule
\multicolumn{4}{c}{\textbf{Time/Energy}} \\
B8 & Lack of time/energy to learn/practice SE [II] & S10 & Quit old job to focus on switching to development [II, III] \\
B9 & Lack of time/energy to find/apply SE jobs [II] &  &\\
B10 & Work responsibilities [I-IV; S10]  &  & \\
B11 & Family responsibilities [I-IV]  &  &\\
\midrule
\multicolumn{4}{c}{\textbf{Psychological}} \\
B12 & Lack of self-confidence [I-III; S12,14]  & S11 & Persevering and not being afraid to try [I-III] \\
B13 & Perception of a low chance of success [I-III; S12-14] & S12 & Increasing motivation with others' success stories [I-III]\\
&& S13 & Having a better understanding of personal goals and values [I, II] \\
\midrule
\multicolumn{4}{c}{\textbf{Other}} \\
B14 & Age-related barriers [I] & S14 & Attending programming meetups [I-III] \\
B15 & Uncertainty due to conflicting opinions [I, II]  & S15 & Going to career fairs [II, III] \\
B16 & Lack of professional networking opportunities [I, III; S14-15] & S16 & Preparing for job interviews [III] \\
B17 & Difficulties finding a job [III; S16-19] & S17 & Building a portfolio of personal projects [II, III] \\
B18 & Difficulty competing/keeping up with colleagues even after & S18 & Improving resume [III] \\
& getting a development job [IV]  & S19 & Applying for jobs that match personal interests [III]\\      
\bottomrule
\end{tabular}
\end{table*}

\subsubsection{RQ1: Barriers}\label{sec:resultsbarriers}
We present the list of barriers
found from our STGT-based analysis process, \review{categorized by its type} (Table \ref{tab:barriersNstrats}). 
The transition process phase(s), presented in Section \ref{sec:model}, that each barrier is applicable to is also placed in brackets following the barrier's name. 
\review{Each barrier is also indexed, e.g., B3.}
For brevity purposes, by default, the pronouns ``they/them'' refer to individuals with non-traditional educational or occupational backgrounds.

\vspace{5mm}
\large
\noindent \textbf{Educational}
\normalsize

\noindent \textbf{B1 Restricted access to desired educational resources [II]} Sometimes, non-financial reasons result in inaccessibility to certain education paths. For instance, someone can afford an undergraduate SE program, but lives far away from a university. There were also cases of an inability to register in SE-related graduate programs due to not meeting the prerequisites and having no alternative methods to fulfill them (e.g., some schools disallow applicants without a SE-related bachelor's degree to become a SE-related graduate student by taking supplementary SE-related undergraduate courses).

\noindent \textbf{B2 Poor quality of educational resources [II]} 
There exists a variety of SE-related educational resources, but their qualities vary. 
There are two relevant aspects: i) what are the criteria when assessing an educational resource's quality, and ii) how can someone know of a resource's lack of quality before committing to it? For the former aspect, relevant criteria varies both between educational paths, and between individuals. For example, both universities and bootcamps might be evaluated by their graduates' employment rates, but bootcamps might also be judged on how practical the skills taught are in real SE jobs. Individual differences matter as well, e.g., some bootcamp students might value the industry networking opportunities that bootcamps offer more than others. 
\review{For the latter aspect, a common method is to search for online reviews from past students. However, we observed individuals wanting to switch into SE commonly reporting difficulties related to finding reviews or discussions on a specific resource (e.g., a less well-known bootcamp), resulting in posts asking for opinions on the quality of specific resources. Unfortunately, sometimes, even these posts do not get many responses.}

\noindent \textbf{B3 Difficulties with comprehending learning materials [II] } 
For individuals with non-traditional educational or occupational backgrounds, learning SE-related concepts could be especially hard, since the types of skills and knowledge familiar to them can be quite distinct from those in SE. For instance, we observed people trying to switch into SE who are professional cooks or nurses, who might have less theoretical background knowledge than someone from fields that are slightly closer to SE, like a physicist. 
As such, even understanding basic computer science topics could be challenging, particularly among those who decided to learn SE through the path of self-learning, since it is often less structured. 

\noindent \textbf{B4 Uncertainty planning an educational pathway [II]} 
One of the most common types of questions we observed from our data is on which educational pathway to choose. This is a complicated type of question for several reasons. First, individual contexts have significant relevance. As discussed by other barriers, factors like whether they have a family, their financial resource, and learning preferences all matter. Second, they may not be aware of all possible educational pathways, and their respective pros and cons. While the more common pathways are more well-known (e.g., in-person university), others might be less known (e.g., free university courses, online universities). Third, there is a balancing act between investing resources into an educational pathway (e.g., paying for a bootcamp) and the pathway's effectiveness towards landing a SE job. All these considerations result in a complicated decision matrix of various possible pathways that vary widely across individuals.

\noindent \textbf{B5 Uncertainty with choosing educational resources (e.g., unsure of what to learn) [II]} 
For those who have chosen a certain educational pathway, another common area of uncertainty is what topics to learn. This is especially prevalent for those who chose to self-learn. Questions on whether they should focus on learning various aspects of software development---mobile application development, front-end programming, back-end programming, etc. are commonplace. For those who chose more structured educational pathways (e.g., online universities), these questions might also appear when they are given the opportunity to select courses that focus on different aspects of programming.

\vspace{5mm}
\large
\noindent \textbf{Financial}
\normalsize

\noindent \textbf{B6 Financial cost of software engineering education [II]} Some SE education options cost more than others, for instance, a full SE bachelor's degree will probably cost significantly more than self-learning. However, financial costs might also vary widely between alternatives of any one path, e.g., degrees from private institutions might cost more than public institutions, and a prestigious programming bootcamp might cost more than other bootcamps. During self-learning, one could unnecessarily spend on paid resources that cover topics also taught by free resources.

\noindent \textbf{B7 Financial cost related to resources for job search [III]} 
During our analysis, we encountered discussions on whether resume-writing/improvement services are worth paying for. Other paid resources that could facilitate the job search process include paid job boards, and paid services for interview practices, like access to coding interview practice questions.

\vspace{5mm}
\large
\noindent \textbf{Time/Energy}
\normalsize

\review{Two frequently discussed barriers are the \textbf{(B8) the lack of time/energy to learn/practice SE [II]} or to \textbf{(B9) find/apply to SE jobs [III]}. The two barriers below are common reasons behind \textbf{B8} and \textbf{B9}.}

\noindent \textbf{B10 Work responsibilities [I, II, III, IV]} Some cannot afford to fully commit to learning and practicing SE. Instead, they have to keep working their current job while learning SE after work. For some, this, in addition to other barriers, like their older age, make them wonder if they could ever ``catch up.'' 
\review{Furthermore, some individuals might choose to accept non-SE jobs that might include some SE-related tasks (e.g., some Business Analysts roles) so that they can better fund their process of eventually getting a proper SE job. However, we observed some instances where this plan backfired; they ended up with even less time to learn and practice SE and reported feeling stuck.}

\noindent \textbf{B11 Family responsibilities [I, II, III, IV]} Some also have to take into account their family responsibilities. A common example is the need for a continuous income source to feed their family while trying to switch into SE. As such, individuals with family responsibilities might be more constrained. For instance, they might be incapable of resigning from their old job to focus solely on learning SE by registering in a SE undergraduate program. Sometimes, there lies a tension between self-interest and family needs. How far should an individual go in their pursuit of a desired career in SE while not sacrificing their ability to meet their family needs?
However, there are also cases where both ends align: successfully switching into SE could yield a career that is more satisfying and affords a more comfortable family life than their current career.

\vspace{5mm}
\large
\noindent \textbf{Psychological}
\normalsize

\noindent \textbf{B12 Lack of self-confidence [I, II, III]} Self-confidence is crucial for career success \cite{hay2006exploring}. However, a lack of self-confidence is not uncommon among those with non-traditional educational or occupational backgrounds trying to switch into SE. While they do not explicitly report lacking self-confidence, we observed reports of feeling stupid and hopeless, that they are only capable of getting horrible jobs, and for those who have been trying to switch for some time, feeling depressed and incapable. 

\noindent \textbf{B13 Perception of a low chance of success [I, II, III]} Those with non-traditional educational or occupational backgrounds might have a perception that they have a significantly smaller chance of succeeding at obtaining a SE career compared to their counterparts with traditional educational or occupational backgrounds. 
The most common examples include the perception that not having a STEM university education or not having prior SE work experience means very low employability within the SE industry.
While these perceptions are partially based in reality, they are often exaggerated enough to make individuals question if it is even worth to try switching into a SE career, i.e., instead of seeing barriers as surmountable over time, barriers are seen as permanently removing any possibilities of successfully switching into SE. 
\review{As such, this barrier is a culmination of the expected barriers one might face if they choose to transition, an assessment of the significance of any expected barrier, and possible negative biases that might exaggerate these assessments.}

\vspace{5mm}
\large
\noindent \textbf{Others}
\normalsize

\noindent \textbf{B14 Age-related barriers [I]} A sizeable part of the population studied by this work has spent a long time (years or even decades) in non-SE education or careers. As such, they are often older than fresh graduates who are also trying to enter the SE job market. It is not uncommon, among analyzed posts, for posters to question if they could switch into a SE career since they are ``old.'' It is clear that this perception of being old is subjective. While it is true that objectively, they are often older than their fresh graduates' counterparts, some saw themselves as ``old'' since they were in their mid-40s, while others perceived themselves as ``old'' in their late-20s, owing to the perception that those entering the SE job market are often in their early-20s. As a result of this age gap, they sometimes perceive themselves as being less cognitively capable and as such, less capable of being competent programmers, often ignoring the complex nature of SE jobs, which commonly involve components that are not purely cognitive in nature, like teamwork and communication.

\noindent \textbf{B15 Uncertainty due to conflicting opinions [I, II]} Another barrier faced by them is the existence of conflicting opinions from those perceived as having more expertise in the SE industry than themselves. From our analysis, there is an implicit expectation for fellow Redditors to provide straightforward answers to questions they have. 
However, the transition process is filled with nuances, resulting in conflicting opinions shared by others sometimes.
This is made worse by the tendency of some Redditors to provide opinions without sufficient explanations for those without a deep understanding of the SE industry.
These conflicting opinions in addition to a lack of understanding about certain nuances result in ineffectiveness when making decisions.

\noindent \textbf{B16 Lack of professional networking opportunities [I, III]} How many professional networking opportunities someone has depends on how they decide to switch into SE and their individual contexts. For instance, attending online university programs might yield less networking opportunities than in-person alternatives. Living in smaller cities might also make it harder to network with other SE.
These networking opportunities could influence whether they decide to switch into SE and how easy it is to get a SE job.

\noindent \textbf{B17 Difficulties finding a job [III]} While job-finding difficulties are common for any jobseeker, we observed those with non-traditional educational or occupational backgrounds to commonly be more clueless about the reasons for their job-seeking difficulties. A few reasons for this were observed. First, some individuals fail to realize that assumptions they hold about how to successfully find a job in a non-SE field that they are familiar with are not true within SE contexts. A common example of this is trending within SE for how resumes should be formatted and presented, and what weight a piece of information holds in their resume. Another observed reason is the failure to accurately assess how ready they are to apply for jobs. For instance, a lack of knowledge of how important SE projects are to job search success might result in an individual perceiving themselves as ready for a job search too early, i.e., before they spend the time to build up an impressive list of SE projects on their resume. Moreover, we observed some cases where people get ``tricked'' into accepting a job that promised software development tasks but ended up not having any SE-related tasks. This could be due to a lack of understanding of how SE jobs are usually structured, and a sense of desperation for a stepping stone. 

\noindent \textbf{B18 Difficulty competing/keeping up with colleagues even after getting a development job [IV]} Sometimes, those with non-traditional backgrounds who successfully landed a SE job might return to the Preparation to change stage in the form of additional learning. Commonly, this entails learning more theoretical SE or CS knowledge, since they only focused mostly on the more practical side of programming in the past. 

\subsubsection{RQ2: Mitigation Strategies}
We present the main list of mitigation strategies that we found from our STGT-based analysis process below, in a similar vein to how the barriers have been presented above (Table \ref{tab:barriersNstrats}). For each of the five types of strategy, we also discuss associations between those strategies and barriers above.

\vspace{5mm}
\large
\noindent \textbf{Educational}
\normalsize

\noindent \textbf{S1 Attending university [II]} \review{SE-related university programs come in various types. Primarily, programs can be categorized in two dimensions: whether they are at the undergraduate or graduate level, and whether they are offered in-person or online, totalling four possible paths. We discuss the nuances below:
\begin{enumerate}
    \item Undergraduate vs. graduate programs: Attending an undergraduate program in a SE-related field is a frequently recommended educational pathway. This is because it has several advantages, including the instruction of crucial theoretical CS concepts, a well-defined and structured multi-year curriculum, and possible opportunities to specialize in specific areas of SE. However, for some individuals aiming to transition into SE, an undergraduate degree alone may not be sufficient, particularly in specialized areas such as artificial intelligence or machine learning where graduate degrees are frequently required. However, this path can be challenging. Some individuals may find themselves ineligible to apply for SE-related graduate programs due to specific admission requirements. Others may face the additional burden of taking supplementary courses to meet the program's prerequisites, further adding to the workload of an already demanding curriculum. Another dimension that adds to this path's complexity is the decision between thesis-based and course-based graduate programs. From our analysis, it seems like having research experience is only valuable for SE jobs that involve research as well. All in all, enrolling in a graduate program is only advisable for those who have a clear goal in mind of wanting to switch to particular sectors of SE where graduate degrees add value, i.e., where research experience or graduate-level computer science concepts are useful. 
    \item In-person vs. online programs: Online programs could provide more geographical and temporal flexibility, and possible financial savings (e.g., reduced accommodation costs). Moreover, from our analysis, the wisdom of the crowd seems to suggest that degrees from online programs are commonly not perceived as inferior by hiring managers when compared to in-person programs. However, in-person programs offer additional benefits in the form of more direct access to teaching assistants for additional support, and in-person interactions with peers and instructors. Furthermore, options of online degrees are often more limited than in-person alternatives. For example, online graduate programs are often only available for course-based, but not thesis-based, options. 
\end{enumerate}
\noindent The four sub-strategies are indexed as:
\begin{enumerate}
    \item \textbf{S1.1} In-person undergraduate programs
    \item \textbf{S1.2} In-person graduate programs
    \item \textbf{S1.3} Online undergraduate programs
    \item \textbf{S1.4} Online graduate programs
\end{enumerate}}

\noindent \textbf{S2 Attending a bootcamp [II]} Bootcamps are educational programs that usually have a very sharp focus on teaching only the necessary practical skills needed for junior SE jobs. While this might seem like the most straightforward way to switch into SE, there are many factors to consider. Particularly, the difference in quality between bootcamps that are perceived as being good-quality vs. lacking quality (based on comments from those who claimed to be ex-students, teaching assistants, or even recruiters) seems to be relatively large. The best bootcamps are often significantly more costly (even though some of them offer creative payment options that reduce the upfront costs), and more intense, involving weeks or months of continuously being taught large amounts of knowledge. However, these bootcamps also seem to offer many useful resources, from instructional support that are highly available, to career advisors with wide networks in the SE industry. In fact, it is not uncommon for these bootcamps to offer some form of employment guarantee. On the other extreme, there are bootcamps that students might describe as scams, ones that teach outdated content using teachers that were not properly vetted for teaching expertise. Regrettably, our research revealed discussions regarding certain bootcamps that employ false advertisements that lure individuals with untrue promises. Individuals with non-traditional educational or occupational backgrounds appear to be more prone to these baits since they have fewer insights into the SE industry, and are hence less able to spot the ``red flags'' of a bad bootcamp, e.g., the syllabus containing many outdated and irrelevant topics. Through our analysis, we found unfortunate examples of individuals giving up on switching into SE after committing all their available financial resources to bad-quality bootcamps.

\noindent \textbf{S3 Attending community college [II]} 
Sometimes, attending a SE-related program in a local community college is suggested, especially if someone did not do very well in high-school. Doing well in community college and earning relevant certificates or diplomas could open up opportunities to transfer into a university SE/CS program, and when compared to university programs, attending community college is a less expensive option to get a better understanding of whether they want to pursue a SE career. Just like university programs, there are both \textbf{(S3.1) in-person} and \textbf{(S3.2) online} options.

\noindent \textbf{S4 Self-learning (e.g., YouTube, Coursera, forums) [II]} On a high level, self-learning is the educational path that is perceived as the most flexible. As long as an individual has access to the internet and a decent computer, they can learn using online resources anywhere, anytime. Moreover, many quality educational resources are available for free. However, self-learning has its own set of weaknesses. Particularly, how effective self-learning depends very much on how much clarity an individual has on \textit{when} to use \textit{which} piece of educational resource; in other words, a roadmap of when to learn what concepts, and how to learn them.
For individuals with non-traditional educational or occupational backgrounds who know next to nothing about SE, their main strategy for figuring out such a roadmap might involve browsing online forums for common suggestions and assessing the suggestions through trial and error. 
Unfortunately, this process could be a huge time sink and demotivate people as a result of gaining little progress despite significant time investments.
That said, for those who need to work a full-time job while learning SE, self-learning might be the only viable educational path.
And for others, self-learning is often an important complementary educational path, since it could be useful for acquiring important SE knowledge not taught in their primary educational path. For example, someone attending university might use self-learning to get educated on practical knowledge (e.g., using GitHub) when their university courses are focusing on more theoretical concepts.
\review{This strategy includes two sub-strategies. Firstly, for those unsure about which software engineering (SE) type to start with, a common suggestion is to \textbf{(S4.1) begin by learning web development}. Reasons include the availability of beginner-friendly project guides online, a perceived larger job market for web development, and lower entry barriers for junior web development roles. Another piece of advice is to \textbf{(S4.2) avoid being overwhelmed by details during self-learning}. To address this, recommendations include staying focused on learning-by-doing projects, acquiring enough knowledge to execute projects without becoming overly ambitious, and seeking beginner-friendly educational resources, possibly those recommended by others on Reddit.}

\noindent \textbf{S5 Finding tutoring [II]} Tutoring has also been suggested as a way to get more personalized help when learning SE. This could be especially useful for individuals who have little to no understanding of what switching to SE and learning SE entails. This strategy's effectiveness depends a lot on the tutor's quality, which could be costly, or less available in less populated areas.

\noindent \textbf{S6 Getting professional certificates [II]} Most of the time, professional certificates are perceived as being more relevant for IT, but not necessarily SE. However, professional certificates could be useful in some cases, like certificates offered by Microsoft that focus on programming.

\noindent \textbf{S7 Choosing educational resources with better personal fit [II]} 
Assessing how each educational path relate to personal preferences could be helpful for making better decisions when choosing an educational path to learn SE. E.g., an individual who knows they do not absorb information well in lecture-style classes might prefer paths and resources that are not delivered through lectures. Similarly, an individual who learns better through more frequent practice should incorporate more practices or projects when learning new concepts.
Using resources that fit better might result in higher learning effectiveness and an easier time learning SE concepts.

\noindent \textbf{S8 Choosing a reputable education path [II]} Reputation could play an important role when selecting an educational path. For example, we found discussions of Redditors debating which particular university had a better reputation in the eyes of the employers, i.e., if graduates from a school are seen as more capable software engineers.
As such, these discussions are often specific to certain geographical locations.
That said, some Redditors warned that reputation should not take priority over fit with personal preferences when choosing educational paths and resources, since reputation would not matter if the chosen educational path is ineffective due to bad personal fit.

\noindent \textbf{Related Barriers} Uncertainties with choosing educational paths (B4) due to a lack of information could be mitigated by being educated on the various possible paths (S1-6) and how they are perceived by employers (S8). Similarly, being informed regarding these paths (S1-6) can help someone find alternatives to educational resources that cannot be accessed (B1), lack quality (B2), or are hard to comprehend (B3). For example, someone struggling to understand a SE-related lecture in university (S1) can use self-learning (S4) to mitigate that.
On the other hand, understanding personal fit (S7) could be important to better select educational paths (B4) and resources (B5).

\vspace{5mm}
\large
\noindent \textbf{Financial}
\normalsize

\noindent \textbf{S9 Getting financial aid [II]} 
Financial costs related to education and job preparation could be high (B6,7), and financial support is a barrier commonly mentioned. 
Suggestions to overcome it refer to getting aid from family members or friends. For veterans in the US, suggestions are to use educational assistance provided by the GI Bill~\footnote{https://www.va.gov/education/about-gi-bill-benefits/} or check for scholarships such as Yellow Ribbon~\footnote{https://www.va.gov/education/about-gi-bill-benefits/post-9-11/yellow-ribbon-program/}.

\vspace{5mm}
\large
\noindent \textbf{Time/Energy}
\normalsize

\noindent \textbf{S10 Quit old job to focus on switching to development [II, III]} Learning and practicing SE, and looking for SE jobs all while working a full-time job (B10) is no easy feat. As such, one possible mitigation strategy is to quit their old job to focus solely on switching to SE. For those with sufficient financial resources, this could be a suitable strategy. Of course, what constitutes sufficient is subjective. We observed, during our analysis, questions on how many months of living expenses individuals should save up in advance before quitting their job to prepare for switching into SE. However, for others, this is overly risky, especially if they have financial dependents.

\vspace{5mm}
\large
\noindent \textbf{Psychological}
\normalsize

\noindent \textbf{S11 Persevering and not being afraid to try [I, II, III]} 
In the face of the various barriers, it is common to see reports of individuals feeling incapable or unmotivated to continue on the journey of switching into SE. As such, although perseverance is beneficial in many areas of life, it is especially important in this context and worthy of being a strategy. Not being afraid to try learning new skills (i.e., new programming languages) is also important. 
Having personal strategies to motivate oneself and nurturing an attitude of not being afraid to try can be significantly beneficial when dealing with barriers.

\noindent \textbf{S12 Increasing motivation with others' success stories [I, II, III]} A common trend we observed in our data is the sharing of success stories from those with non-traditional backgrounds who successfully switched into a SE career. These stories inspire others still transitioning. 
This is especially true when an individual's non-traditional educational or occupational background matches that of someone who successfully switched, since the success stories are clear examples that they could succeed despite barriers due to their non-traditional backgrounds. 
As a side note, these stories could also impart important lessons on how to deal with barriers faced when switching.

\noindent \textbf{S13 Having a better understanding of personal goals and values [I, II]} As discussed in the barriers, individuals commonly have a hard time understanding what to eventually focus on when learning SE, and if a SE career would be a good fit for themselves. Towards this, time could be spent reflecting on one's own goals and values. Having better self-understanding could provide clearer insights into the best path to take toward a SE career.

\noindent \textbf{Related Barriers} Understanding others' experiences (S12) could help boost self-confidence (B12) and have a more positive outlook on one's chances of success (B13), while a deeper understanding of one's own goals (S13) can be a good first step towards understanding the reasons behind perceptions of having a low chance of success (B13).

\vspace{5mm}
\large
\noindent \textbf{Others}
\normalsize

\noindent \textbf{S14 Attending programming meetups [I, II, III]} Programming meetups are good not just for job searching, but also for getting advice on switching into SE and the local SE industry. 

\noindent \textbf{S15 Going to career fairs [II, III]} Career fairs offer great opportunities for those trying to switch into SE to i) network with hiring managers and other jobseekers, ii) gain insights into strategies for switching into SE through communication with other attendees (phase III), and iii) have a better understanding of what kinds of skills and experiences are sought after by hiring companies for specific types of SE jobs (phase II and III). 

\noindent \textbf{S16 Preparing for job interviews [III]}
Preparing for SE job interviews could be quite different from job interview preparations in the fields that those with non-traditional educational or occupational backgrounds are familiar with. 
The main task for this preparation involves practicing coding questions through platforms like \textit{LeetCode}. 

\noindent \textbf{S17 Building a portfolio of personal projects [II, III]} Comments from our analyzed data often emphasized the importance of building up a portfolio of personal SE projects. 
Benefits include tangibly demonstrating SE skills, practicing SE skills such as teamwork skills needed to contribute to open-source projects, and exploring what aspects of SE they are most interested in, e.g., if someone does not enjoy doing a  front-end web project, it could be a good signal that they should try other kinds of projects.

\noindent \textbf{S18 Improving resume [III]} Having a good-quality resume is essential for success in the job search phase. As such, various ways to improve one's resume should be employed, especially since having non-traditional educational or occupational backgrounds might mean not knowing what constitutes a ``good'' resume. These ways could include paying for resume improvement services and gathering professional software engineers' feedback on their resumes through various avenues (e.g., online forums, and in-person gatherings).

\noindent \textbf{S19 Applying for jobs that match personal interests [III]} Other than considering how much a job pays, Redditors also advised people to also consider how a job could pave the way towards a long-term career that matches their own personal interests when applying for jobs.

\noindent \textbf{Related Barriers} Similar to learning about others' success stories (S12), attending developer meetups (S14) and talking with other developers, potentially some of whom with non-traditional backgrounds as well, could help mitigate psychological barriers (B12,13). Moreover, meetups (S14) and career fairs (S15) could provide networking opportunities during job search (B16), while the various job search strategies (S16-19) could mitigate difficulties during job search (B17).

\subsection{Follow-up Survey}

\subsubsection{Participants}\label{sec:participants}

A total of 55 participants indicated that they fulfilled the survey's inclusion criteria and gave their consent. However, six responses were deemed invalid. Two responses were entirely empty, while four responses only provided demographic information but did not respond to any questions on personal experiences, barriers, or mitigation strategies. Additionally, three more responses were categorized as invalid since, despite self-reporting as meeting the inclusion criteria, these participants reported not having any non-traditional educational or occupational backgrounds in the survey (e.g., reporting having a bachelor's degree in CS and working in the SE field, without any experience in other fields). Consequently, the analyses were conducted on 46 responses.

Thirty-three participants self-identified as men, 12 as women, and one preferred not to say. Thirteen participants identified as 18-25, five as 26-30, 12 as 31-35, six as 36-40, three as 41-45, three as 46-50, three as 51-55, and one as over 65. 
In terms of racial-ethnic background, 18 participants identified as White or Caucasian, 16 as Asian, 10 as Latino or Hispanic, three as Black or African American, three as Middle Eastern, and one as Indigenous.
Participants also reported a variety of native languages, including Portuguese (n=14), English (n=12), Chinese (n=4), French (n=3), Cantonese (n=2), Hindi (n=2), Bengali (n=3), and n=1 for Italian, Arabic, Dutch, German, Farsi, Zulu, Tamil, Ukrainian, Urdu, and Yoruba.

Participants reported prior experiences in various non-SE fields, from finance to the arts (Table \ref{tab:nonseexp}). 
\review{All participants, except P18, possess some type of post-secondary credential. Sixteen participants have either obtained, or are completing, some form of post-secondary program in SE/CS. Further details on their post-secondary education backgrounds can be found on Zenodo.~\footnote{Please refer to \url{https://zenodo.org/records/10511167}}}
Moreover, participants have/had a variety of software-related jobs (Table \ref{tab:sejobs}). The high amount of job entries (115 in total for 49 participants) might be indicative of the SE industry's relatively mobile nature \cite{naresh2015job,fallick2006job}.

\begin{table*}[h]
\caption{Participants' non-SE educational and occupational backgrounds.}\label{tab:nonseexp}
\begin{tabular}{@{}llll@{}}
\toprule
ID & Edu./Occ. Background  & ID & Edu./Occ. Background\\
\midrule
1 & Banking \& Finance & 24 & Global Business \& Digital Arts \\
2 & Marketing \& Design & 25 & Accounting \& Auditing \\
3 & Mechatronics Engineering & 26 & Mathematics \\
4 & Law \& Finance & 27 & Mathematics \\
5 & Architecture & 28 & Psychology \\
6 & Physical Education \& Nursing & 29 & Mathematics\\
7 & Electronics \& Instrumentation & 30 & Economics \& Biblical Studies\\
8 & Electrical Engineering & 31 & Copywriting \& Graphic Design \\
9 & Finance & 32 & Mechatronics Engineering \\
10 & Economic & 33 & History\\
11 & Civil Engineering & 34 & Technical Theater\\
12 & Education & 35 & International Relations \& Agribusiness \\
13 & Nutrition & 36 & Finance\\
14 & Aerospace Engineering & 37 & Business\\
15 & Telecommunication Engineering & 38 & Physics\\
16 & Film \& TV & 39 & Economics\\
17 & Control \& Automation  & 40 & Marketing \& Advertising\\
&  Engineering &  & \\
18 & Accounting & 41 & Education \& Aviation\\
19 & Computer Engineering \&  & 42 &  Biotechnology\\
 & Computational Mathematics & &\\
20 & Chemistry & 43 & Business Administration\\
21 & Banking \& Customer Service & 44 & Electrical Engineering \\
22 & Hardware Verification & 45 & Actuarial Science\\
23 & Education & 46 & Education \& Accessibility\\
\bottomrule
\end{tabular}
\end{table*}

\begin{table*}[h]
\caption{Participants' past and current software related jobs.}\label{tab:sejobs}
\begin{tabular}{@{}llll@{}}
\toprule
Job Title & \#  & Job Title & \#\\
\midrule
Software developer/engineer  & 20 & Game developer  & 4 \\
Front-end web developer  & 16 &  Quality assurance engineer & 3    \\
Full-stack web developer  & 11    & Cloud architect/engineer    & 2 \\
Back-end web developer  & 11   & Data science\footnotemark[1] & 1   \\
Backend engineer  & 9 & Data analyst\footnotemark[1] & 1   \\
AI programmer  & 9  & Business intelligence report developer\footnotemark[1] & 1 \\
Mobile applications developer & 8  & Graphics programmer  & 1 \\
Automation engineer - software & 6  & Software development researcher\footnotemark[1]  & 1\\
Application architect/engineer & 5  & Graduate student\footnotemark[1] & 1 \\
DevOps engineer & 5 & & \\
\bottomrule
\end{tabular}\\
\textit{[1] Free text response from the `Other' option.}
\end{table*}

\begin{figure*}
\centering
\begin{minipage}{.8\textwidth}
  \centering
  \includegraphics[width=.9\linewidth]{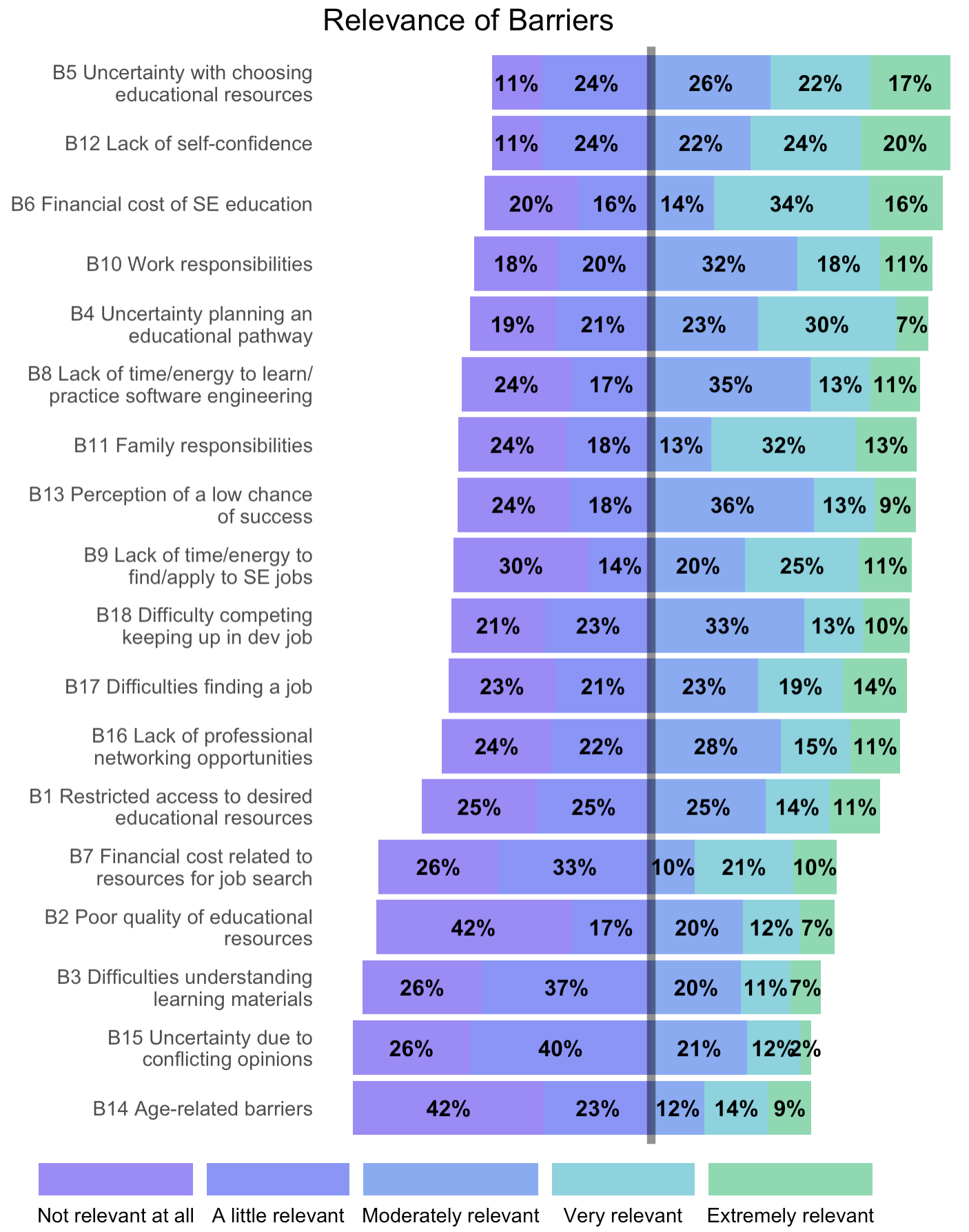}
  \captionof{figure}{Relevance of barriers from the questionnaire.}
  \label{fig:barriers}
\end{minipage}%
\end{figure*}

\begin{figure*}
\centering
\begin{minipage}{.9\textwidth}
  \centering
  \includegraphics[width=.9\linewidth]{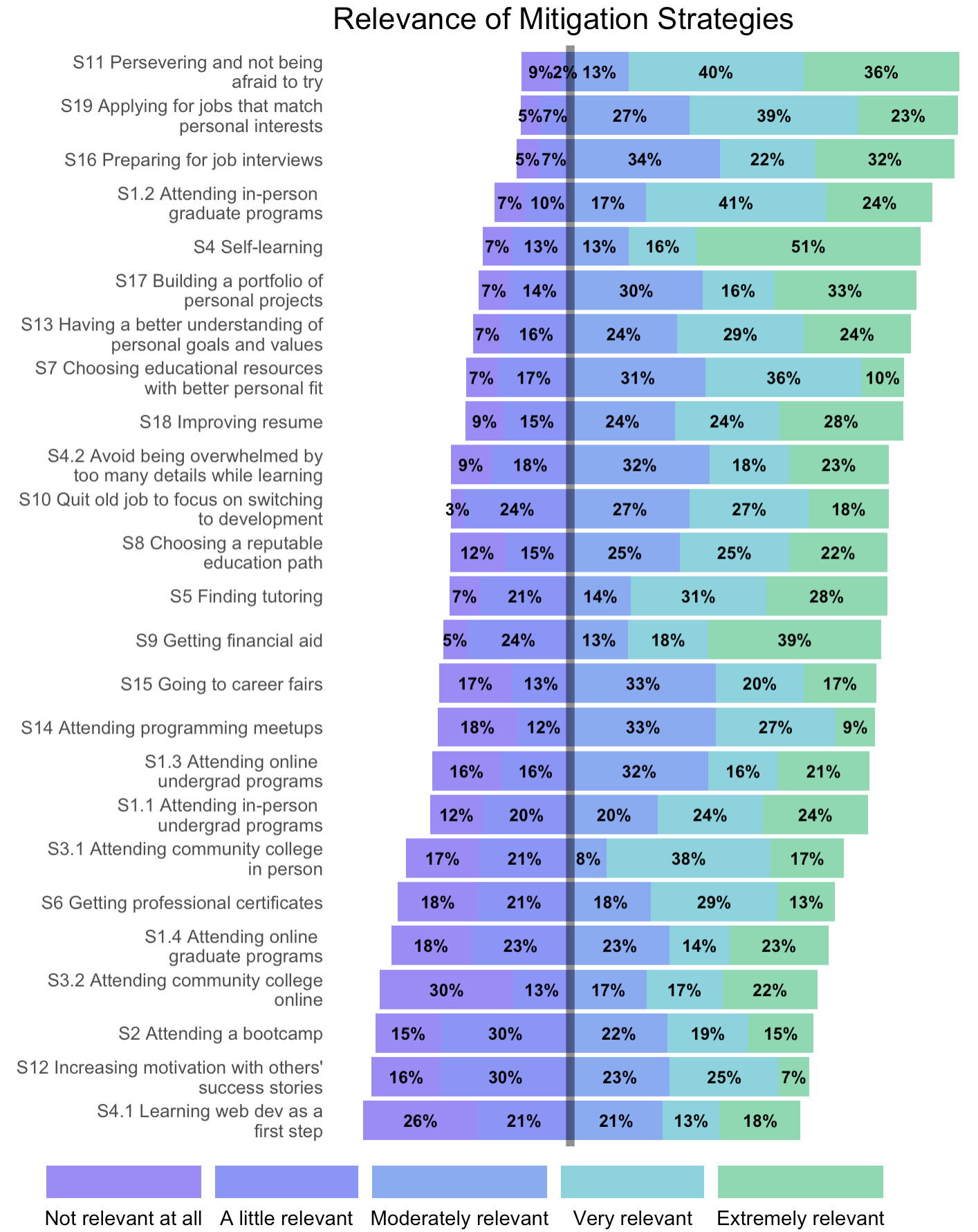}
  \captionof{figure}{Relevance of mitigation strategies from the questionnaire.}
  \label{fig:ms}
\end{minipage}
\end{figure*}

\subsubsection{RQ1: Barriers}\label{results:surveybarrier}

Fig.~\ref{fig:barriers} shows participants' relevance ratings of the barriers from the exploratory study. We consider ``Moderately relevant'' as a positive response instead of a neutral response, hence its position at the right of the vertical line in the chart.
Barriers varied considerably in terms of rated relevance, with barriers like feeling uncertain with choosing educational resources and a lack of self-confidence being rated as at least moderately relevant by around 66\% of participants, in contrast to 35\% for age-related barriers.
Participants were also provided the opportunity to report other barriers not in the list through free-form text responses. A few participants provided responses related to the barriers in the list. For instance, a few participants reported psychological barriers, including ``imposter syndrome'' (P5), ``fear of disappointment'' (P45).
P12 reported difficulties with understanding ``the foundational [way of] thinking to solving problems'' initially as a self-taught SE. P15 reported having ``many resources, but [found it] difficult to start with something'', another self-taught SE, P20, reported having had ``gaps in [their] knowledge when faced with things that most Computer Science students learn quite early on, like common algorithms.'' Besides, P44 reported a barrier due to the ``long and difficult technical interview process'', and P46 reported difficulties with ``working and studying part-time'' without further explanations. Lastly, P40 reported facing difficulties with job search in Canada due to having ``no Canadian experience'', despite prior experiences overseas.
Participants also reported intersectional barriers, including:
\begin{enumerate}
    \item \textbf{Language barriers}: A few participants reported barriers associated with language, since ``most of [the] good content is in English (which is not my native language)'' (P2). P25 also reported similar sentiments, and gave examples: ``Medium, Udemy, EDX, etc.''
    \item \textbf{Gender barriers}: P5, who self-identified as a woman, ``felt a gender barrier''. P40 also reported facing cases of misogyny, ``everyone is keeping an eye on me to get knocked up.''
    \item \textbf{Financial barriers}: These could include financial resources related to the ``cost of living during the study period since most people can't afford [the] cost of living without work[ing]'' (P4), and to ``purchase equipment [for] work'' (P2).
\end{enumerate}

\subsubsection{RQ2: Mitigation Strategies}

We asked participants to rate how relevant each applicable mitigation strategy is/was at mitigating barriers while switching.~\footnote{\review{Sub-strategies were listed as separate strategies in the survey to get more granular results.}}
Figure \ref{fig:ms} shows that for all mitigation strategies, at least 56\% of participants rated them as moderately, or more, relevant while switching to SE. 
\review{Other than that, a few interesting insights could be made. First, overall, participants rated the mitigation strategies as more relevant than the barriers. This could imply that while barriers are usually more specific to each person trying to switch into a SE career, there are relatively more standardized mitigation strategies. A good example of a rather standardized and much recommended mitigation strategy across both studies is the importance of preparing for job interviews by practising coding questions with commonly used platforms like \textit{LeetCode}.
Second, the mitigation strategy rated as the most relevant--persevering and not being afraid to try--was rated as relevant by 89\% of participants, even higher than psychological barriers like a lack of self-confidence (66\%) and a perception of a low chance of success (58\%).
This may hint at the challenging nature of this transition process: even those with no lack of self-confidence and know they have a fair chance at success perceive perseverance and courage as extremely valuable, above other more tangible mitigation strategies.
Another interesting observation is that learning web development as a first step was rated as the least relevant strategy (52\% relevant), and a Mann-Whitney test~\footnote{One of the two groups failed to pass the Shapiro-Wilk’s normality test ($W = .85, p < 0.01)$), hence the use of a Mann-Whitney test instead of an unpaired one-sample T-test.} comparing the ratings of those with (22/46 participants) vs. without (24/46 participants) a web developer job title (i.e., any of Front-end/Full-stack/Back-end web developer in Table \ref{tab:sejobs}) showed that the former group \textit{did not} rate this mitigation strategy significantly more relevant than the latter group ($W = 196, p =.3$). 
In other words, regardless of whether participants has/had job titles with ``web developer'' in them, learning web development as the first step was not very relevant.
}

When asked if there were other mitigation strategies, participants provided the following list:
\begin{enumerate}
    \item Incorporating programming into current job to ``assist [with] current responsibilities'' (P40).
    \item Budgeting for a period of full time study (P35).
    \item Participating in various networking-related strategies such as engaging with online developer communities (P33), attending developer conferences (P36, P41), talking to experienced SEs (P16), and keeping an updated network of SEs (P2).
    \item Learning programming languages that have ``more opportunities in the region'' (P26).
    \item Collaborating and learning with others also switching into SE (P21).
    \item Identifying the correlation between one's previous work experience and their current work in SE (P9).
    \item Doing volunteer work (P48). Although P48 did not provide examples of volunteering avenues, prior works show that volunteering, e.g., participating in open source projects, could be beneficial for a SE career (e.g., \cite{hann2013all}).
\end{enumerate}

\begin{figure*}[t]
    \centering
    \includegraphics[width=.65\linewidth]{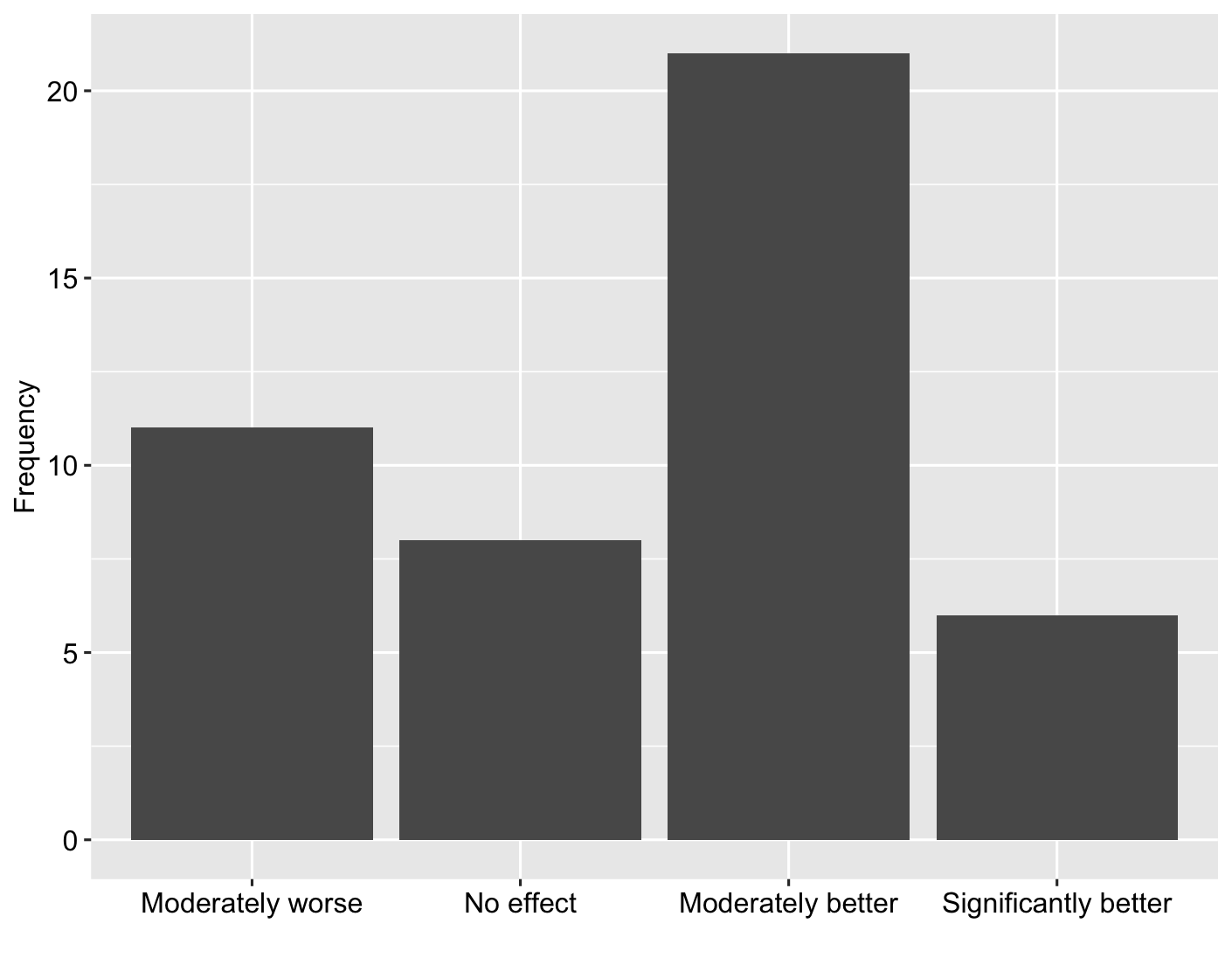}
    \caption{Participants' 5-point Likert-scale (from ``Significantly worse'' to ``Significantly better'') response to the question ``How do you believe your educational/occupational background that is not in software development affected the barriers you faced when switching into software development?'' No participant responded with ``Significantly worse.''}
    \label{fig:effectsnontrad}
\end{figure*}

Many participants provided positive ratings (Fig. \ref{fig:effectsnontrad}) when asked about the impact of their non-SE educational and occupational backgrounds on the barriers they encountered while transitioning into SE; \review{i.e., a rating of ``Significantly better'' would mean that a participant perceived their non-traditional educational/occupational background to have significantly helped them when facing barriers during the transition process}. Additionally, an open-ended question was included to allow participants to provide further elaboration on this aspect.
Here, we present the different types of possible benefits:
\begin{enumerate}
    \item \textbf{Soft skills}: Learning strategies (P20, P43), patience (P43), transferring knowledge to others (P12), communication skills (P6, P16, P31, P33), workplace ethics (P16), project management (P16), teamwork (P16), user/customer experience (P16, P18, P28, P30) and skills associated with the ``de-escalation of crisis situations'' (P6). Without their non-traditional backgrounds, people might ``hardly get'' to learn these soft skills (P1). These soft skills could be useful both during the interview process and to perform tasks in SE jobs (P6).
    \item \textbf{Domain expertise}: Individuals who have successfully transitioned into SE roles and previously worked in companies operating in non-SE fields or with relevant connections to such areas may find their pre-existing domain expertise to be valuable and advantageous. 
    Some provided examples are nutrition (P13), business operations (P37, P39), audit (P25), mechanical (P3) and electrical engineering (P3, P38). Moreover, participants highlighted that prior experiences in other fields offered them a valuable perspective on how software could be effectively utilized (P15,34).
    \item \textbf{Problem-solving}: Participants with previous experiences in fields like architecture~(P5), quantitative data analysis~(P11), or mathematics~(P26, P29) reported being able to transfer skills like problem-solving (P5, P11), flexible thinking (P10), logical (P11, P29) and abstract (P26) reasoning. 
    \item \textbf{Connections}: Individuals who successfully made the switch to software engineering (SE) enjoyed the advantage of having connections with ``two different groups of people: tech and non-tech skilled individuals'' (P2). 
    \item \textbf{Maturity}: Having prior experiences in other fields could, in general, provide people with ``more maturity when switching to a new career'' (P13), and help them ``not make the same mistakes twice'' (P27).
\end{enumerate}

\section{Theoretical Model}\label{sec:model}
\review{

\begin{figure*}[t]
    \centering
    \includegraphics[width=\linewidth]{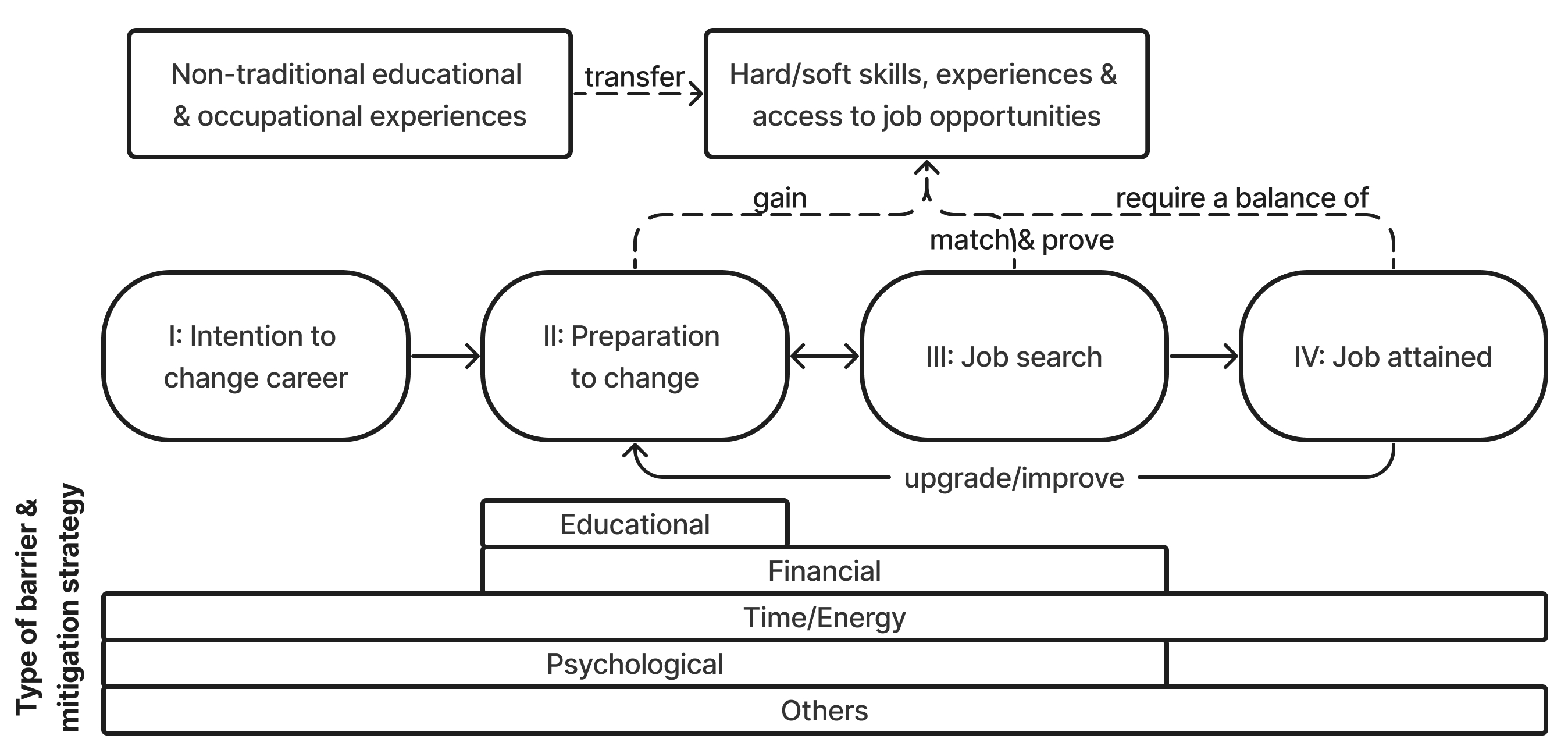}
    \caption{\review{An overview of the proposed theoretical model describing the process of switching to a SE career for those with non-traditional educational and occupational backgrounds, alongside the types of applicable types of barriers and mitigation strategies for each phase.}}
    \label{fig:model}
\end{figure*}

Given the findings in Section \ref{sec:find}, we present an overview of the theoretical model that emerged over the research (Fig. \ref{fig:model}).
This model uses the Integrated Career Change (as discussed earlier in Section \ref{sec:careerswitching}) as a basis.
Importantly, our study does not aim to answer \textit{why} people want to switch to SE,
but people's experiences \textit{when} changing careers to SE. As such, of the 17 components of the Integrated Career Change Model,
the last three are the most relevant: i) intention to change careers, ii) preparation for change, and iii) actual change. 
Interestingly, Rhodes and Doering \cite{Rhodesmodel} noted that preparation for change could occur during the actual change, which they define as beginning ``with leaving the old job and is not completed until the new job/career is entered'', and may not always lead to actual change. 
However, in the context of this work, the aforementioned logic is not always true. Rather, we found it quite common for people to not leave their existing non-SE job till various stages of the change process. As such, we do not see much meaning in defining what constitutes an ``actual'' change. Instead, using these three components as building blocks, we propose a model (Fig. \ref{fig:model}) containing four phases: (I) intention to change careers, (II) preparation to change, (III) job search, and (IV) job attained, which describes the main phases of switching to a SE career. 

\begin{enumerate}
    \item \textbf{(I) Intention to change careers}: Thinking of whether to take the first step towards changing careers to SE is often not an easy or straightforward decision. The severity of existing and potential barriers one might face and the availability and feasibility of mitigation strategies play important roles in explaining the thought process of those who decide to, or give up on, switching into SE.
    \item \textbf{(II) Preparation to change}: This phase refers to the process of gaining SE-related hard skills (e.g., Android programming), soft skills (e.g., working with other developers), and experiences (e.g., of building Android apps). It is an important phase, especially since we are concerned with those with non-traditional occupational or educational backgrounds. Moreover, the methods of preparation can vary widely between individuals, depending on their circumstances. As noted by Rhodes and Doering \cite{Rhodesmodel}, not all who embark on the process of preparing to switch into SE (or job search) successfully get a SE career.
    \item \textbf{(III) Job search}: This is the process of searching and applying for jobs. A smooth job search process often requires successfully proving one's own hard skills, soft skills, and relevant experiences. It also entails having at least a minimum amount of access to SE job opportunities. It is common for someone to go back-and-forth repeatedly between phases II and III, or be in both phases at the same time. E.g., someone might start to apply for jobs and realize they do not know enough, or begin applying for jobs while at the tail-end of their preparation phase.
    \item \textbf{(IV) Job attained}: This phase is when an individual successfully attains, and accepts, offers to SE jobs. However, it is not always where the process ends. As explained in Section \ref{sec:find}, those with non-traditional educational or occupational backgrounds 
    might either take multiple jobs as stepping stones towards a SE job. Even those who attained their first SE jobs might, at some point in time, perform further preparations to upgrade or improve their SE careers. We opted for the job search and attained phases instead of \textit{actual change} in the original Integrated Career Change Model because the definition of \textit{actual change} is too limited for our context. For example, people may or may not decide to leave their old jobs, and those who might do it at different parts of the process, be it before starting their preparation (e.g., having lost their old jobs), or during advanced preparation (e.g., having already built up a sizable amount of SE experience from personal SE projects). 
\end{enumerate}

Additionally, findings from the follow-up survey suggest certain transferable aspects from someone's non-traditional educational or occupational background. These aspects may include hard skills (e.g., logical deduction), soft skills (e.g., socializing skills), relevant experiences (e.g., working with programmers) or access (e.g., existing contacts with knowledge of available SE roles), and this has also been reflected in the model.
Furthermore, several types of barriers and mitigation strategies in each phase were observed during the studies, including educational, financial, time/energy, psychological and other miscellaneous ones.
While certain barriers and strategies are specific to certain phases (e.g., education-related ones during preparation), some are relevant across most or all phases. Recognizing the severity of these barriers and the corresponding strategies in each phase might help with understanding the time individuals spend in each stage. For example, difficulties accessing educational resources may delay the preparation phase, while severe self-confidence issues could impede progress in all phases, leading to career indecisiveness \cite{leong1996construct} (1st phase), reduced learning effectiveness \cite{akbari2020students} (2nd phase), and less successful interviews \cite{dimopolulos2020applicant} (3rd phase).  

As discussed by Corbin and Strauss \cite{corbin2014basics}, unlike descriptions of a phenomenon, a theory is capable of predicting how an individual of the studied population act under certain circumstances. In our context, how one might choose to switch into SE and the outcome of the transition can be predicted by which phase they are in, the barriers faced and the strategies available.
Given an individual's personal circumstances, the model could provide possible insights into the types of barriers and mitigation strategies that an individual might face or have.
To aid with relating each barrier and mitigation strategy observed from the data with the model, the labels of the stages that are relevant for each barrier and strategy have been placed in brackets when presenting the findings from the exploratory study (Section \ref{sec:resultsexplore}).

Based on this model, future work could focus on i) a particular phase of the process (e.g., focusing only on the experience of those who got their very first SE job), ii) a particular type of barrier/strategy (e.g., only exploring psychological processes along the transition), or iii) investigating more thoroughly what aspects are transferable from past experiences.

}

\section{Discussion}
\label{sec:disc}

In this work, we investigate the barriers and possible mitigation strategies of those switching into SE with non-traditional educational and occupational backgrounds. Using a STGT-based method, initial themes and lists of barriers and strategies were analyzed from relevant Reddit data. A follow-up survey was performed to gain a deeper perspective on the relevance of this analysis. Both these studies then contributed towards a proposed model that describes the transition process. Below, we discuss both research questions by comparing findings from both the exploratory study and the follow-up survey, followed by a few other implications.

\subsection{RQ1: Barriers}

\noindent\textbf{Intersectional Barriers}
Since there is very little existing work on the experience of those in SE with non-traditional educational and occupational backgrounds, the intention of this work was to research this in particular, while leaving research on intersectionality with other diversity factors to future work. However, our findings suggest that those with non-traditional education and occupational backgrounds often face barriers that are intersectional with other diversity factors, including language, racio-ethnic background, gender and age. 
This is important, since it implies that the switching process into SE for those with non-traditional educational and occupational backgrounds includes not just barriers that come with having to learn SE and enter its workforce, but also barriers that stem from other diversity factors. 
Moreover, intersectional barriers could manifest both directly and indirectly. 
An example of a direct effect is the relatively little educational SE resources available in less popular languages, which could make the process of learning SE significantly harder, since one might also have to learn a more popular language (e.g., English) simultaneously. 
On the other hand, although not observed in this study per se, indirect effects could also be possible. For instance, those with non-traditional educational and occupational backgrounds that are also ethnic and racial minorities could be in less favorable socioeconomic circumstances with higher probabilities \cite{apasesrace}, which could make significant financial barriers like the expensive cost of SE education, or the budget needed to focus on learning SE without working, even worse. 

Regarding intersectionality, our study observed an interesting pattern with age. In the exploratory study, many were doubtful of a successful switch due to their (relatively older) age. However, age-related barriers were rated as the least relevant by survey participants in the follow-up survey. One possible reason for this discrepancy is that while many of the Reddit posts with age-related concerns were created by those thinking of switching, most of those who participated in the survey have actually made the switch into SE. As such, age-related concerns might be a commonly assumed barrier that is less relevant in actuality. That said, the relevance of barriers might vary across the different SE sectors. For instance, age-related barriers might be more pronounced in SE companies that are more focused on hiring fresh graduates.

\noindent\textbf{Psychological Barriers}
Both the exploratory study findings and the survey results align in terms of how relevant the psychological barriers are. 
Even though the subreddits analyzed in the exploratory study are entirely dedicated to SE-related topics, there were entire posts and numerous comments focused on psychological issues during the transition process. 
Similarly, in the survey, among all the barriers, a lack of self-confidence had the highest proportion of participants giving it an ``Extremely relevant'' rating (20\%) and the second most relevant barrier overall (Fig. \ref{fig:barriers}).
And we posit that the relevance of the barrier of having a perception of a low chance of success would probably be even higher than in this work's survey (58\% of participants rated it at least moderately relevant) if the actual population of SEs with non-traditional backgrounds had a lower proportion of individuals with SE/CS-related undergraduate education than the participants in this work.

Moreover, the data analyzed in the exploratory study is from before the global pandemic happened. 
Not only did the global pandemic negatively affect the mental state of many \cite{bozdaug2021psychological,mukhtar2020psychological}, it also motivated many more individuals to switch to SE \cite{cbccoding}, since the SE industry was impacted less by the pandemic when compared to other industries \cite{arthur2021studying,hylton2022long}.
As such, it is very possible that the severity of the psychological barriers faced by those trying to switch into SE during the pandemic was even worse for reasons like the additional pressure of needing a new job due to financial stress \cite{obrenovic2021threat}.

\noindent\textbf{Unexpectedly Less Relevant Barriers}
Contrary to our expectations, during the follow-up survey, we found that five of the barriers were rated as either ``Not relevant at all'' or ``A little relevant'' by more than half of the participants, despite initially anticipating their relevance.
\begin{enumerate}
    \item \textbf{Financial cost related to resource for job search}: This might be because most people did not invest financially into resources related to the job-search process, and aligns with evidences from both Reddit and survey participants that free SE educational resources are relatively easy to access.
    \item \textbf{Poor quality of educational resources}: Similar to the last one, this could be due to the existence of high quality educational resources for SE that are both public and free online.
    \item \textbf{Difficulties with comprehending learning materials}: It is possible that learners could easily find alternative learning materials that are more suitable to their expertise level, thanks to the abundance of educational resources online. However, while educational resources for all levels of expertise might be accessible, how to select the best resource to learn from is an entirely different topic, and was rated as one of the most relevant barriers.
    \item \textbf{Uncertainty due to conflicting opinions}: It is plausible that this barrier is more prominent among individuals who extensively engage with online SE communities and are exposed to a broader range of diverse opinions concerning the topic of transitioning into SE. Although we suspect that the average level of engagement with online communities among participants may be lower than that of Reddit users, we cannot definitively confirm this as it was not measured in the survey. 
    \item \textbf{Age related barriers}: This is the most surprising result, since our exploratory study seems to suggest a much higher relevance. A simple regression analysis was performed for more insights, but did not yield any statistically significant results. However, it is important to acknowledge that the results of this study alone do not provide sufficient evidence to draw conclusions about the reasons behind this discrepancy in relevance. Future research is warranted to gain a deeper understanding of this phenomenon. 
\end{enumerate}

\noindent\textbf{Coverage} While this work serves as a good first step when it comes to identifying some of the barriers faced by those transitioning into a SE career with non-traditional backgrounds, by no means did this work uncover all of them. 
The categories of educational, financial, time/energy-related, and psychological barriers are also likely to be incomplete. With further research, we posit that new categories will emerge from the ``Others'' category.

\subsection{RQ2: Mitigation Strategies}
Overall, the findings from the survey on the relevance of strategies observed in the explanatory study suggest many strategies that are greatly relevant to the transition process; we discuss these strategies below. Interestingly, the relatively higher relevance of most of the strategies when compared to the barriers seem to suggest that while barriers might vary significantly across individuals, there exists a set of strategies that are more common.

\noindent\textbf{Formal vs. Informal Education} 
The overall sentiment on formal education from our exploratory study was that formal education could be significantly beneficial to landing a SE job, but not necessary. There was also a general sentiment that formal education is simply not for everyone. 
And while the importance of having a diploma or degree on the resume when applying for jobs might be decreasing, the educational benefits of formal education are very dependent on an individual's preferences, traits and the institute and program chosen.
Moreover, both the Reddit data and survey participants share the sentiment that in-person formal education is more beneficial than their online counterparts.
In contrast to the exploratory study, the follow-up survey indicated that participants considered  in-person graduate education more relevant than in-person undergraduate education. This difference might be attributed to  a higher proportion of survey participants having graduate education compared to Redditors whose comments were analyzed in the exploratory study. It is important to acknowledge the potential discrepancy in accurately representing the proportion of software engineers with non-traditional backgrounds and graduate SE/CS degrees in both the Reddit data and the survey participants. Nonetheless, very few Reddit users in our analyzed data reported having graduate education in SE/CS. Moreover, from the exploratory study, graduate education in SE/CS was perceived as highly inaccessible to most people trying to switch to SE due to prerequisite requirements. Therefore, those who successfully enrolled in graduate SE/CS programs, such as some survey participants, may view having a graduate education as particularly relevant.
Furthermore, just like the exploratory study, the follow-up survey results showed that the cost of SE education is quite a significant barrier, one that many might not be privileged enough to overcome. However, there are fewer findings on how various parties can better help individuals overcome this barrier, beyond just trying to get financial aids (S9) somehow.

In terms of learning, the large amount of highly accessible educational SE materials online has increasingly become a viable alternative to getting some form of formal education.  However, it is a double-edged sword, as the variety of online resources could easily overwhelm newcomers. This observation aligns with research on the choice overload effect in information retrieval contexts, i.e., users could get overwhelmed by large amounts of choices \cite{beierle2017exploring}.

\noindent\textbf{Psychological Strategies}
Similar to the psychological barriers, various psychology-related mitigation strategies were rated as highly relevant. Specifically, 89\% of participants rated the strategy of persevering and not being afraid to try as at least moderately relevant, and 77\% for having a better understanding of personal goals and values. 
This aligns with the exploratory study's findings; whenever psychological issues were discussed, Redditors often encouraged each other to persevere.
While most people trying to switch to SE have an awareness of how to access educational resources online, it is unclear i) how accessible various forms of psychological support are to them, and ii) whether they are aware of any available support and are willing to use them.

\noindent\textbf{Learning Web Development} 
Several strategies were consistently salient across both the exploratory study and survey, e.g., building a portfolio of personal projects, preparing for job interviews, and others that were discussed above. 
That said, there were also inconsistencies.
For example, learning web development as a first step when venturing into SE seemed like a relatively unanimous opinion within the analyzed Reddit data when it comes to questions surrounding what subtopics of SE to first learn. 
And the SE job experiences of the survey participants seem to align with this observation: around half of the participants has/had job titles that explicitly include the phrase ``web developer.'' This does not include other job titles (e.g., software developer) that might also include tasks that could fall under the umbrella of web development.
Regardless, it is clear that web development roles are one of the most common in the SE industry.
As such, it seemed reasonable to expect participants to rate the mitigation strategy of learning web development as the first step to be relatively relevant. 
However, it turned out to be the least relevant strategy out of the entire list presented in the survey.
Due to the need to keep the survey relatively short so that the responses do not suffer from decreases in quality due to participant fatigue, it did not include questions on \textit{why} they rated a certain strategy more/less relevant. 
It is hence impossible to really know the true reasons behind why this strategy was perceived as the least relevant. 
However, a possible reason is that while many of them eventually learned about web development, very few of them actually learned it just as they were starting to learn SE.
This aligns well for the around one-third of participants who attended/are attending a SE/CS-related program in a post-secondary institution~\footnote{Please refer to \url{https://zenodo.org/records/10511167}}, since these programs might rarely teach web development directly in their CS1 courses (i.e., introductory CS courses).
For those without post-secondary education in SE/CS, it would be interesting for future works to understand what this population's education path is like, and where exposure to web development happens in that learning process.
New strategies could be created if such research observes benefits associated with earlier exposure to web development while transitioning into SE.

\noindent\textbf{Were All Barriers Mitigated?}
Unfortunately, not all barriers that were identified in the exploratory study had corresponding mitigation strategies from either the exploratory study or the survey.
For instance, there were no strategies to mitigate the lack of time/energy to learn/practice SE (B8) and apply to jobs (B9). And while general time management strategies could be useful, future works could investigate specific time management methods used while learning SE or applying to SE jobs. An example here could be ways of making the transition process more efficient; it could be learning to identify how useful a piece of educational resource is so that no additional time is spent on less useful ones, or tools to make the job application process more convenient. 
Our work also did not identify strategies to better navigate family responsibilities (B11). This is unsurprising, since that seems like a topic more general than getting a SE career per se; we discuss the generalizability of these findings later in Section \ref{sec:discgeneralize}.
Furthermore, while having a better understanding of personal goals (S13) could be helpful to mitigate perceptions of a low chance of success (B13), it is likely to be insufficient; the ability to be realistic, adjust these goals and generate good plans are examples of complementary strategies.
More relevantly, future SE research could explore other strategies aimed at helping SEs with non-traditional backgrounds perform at their best in the workplace (i.e., phase IV in our proposed model). E.g., do technology companies offer mentorship or funding opportunities for continuing education aimed at non-traditional SEs?
Overall, findings from this work are, in a way, observations of the most commonly discussed mitigation strategies online. And while the survey found high levels of relevance, this work opens many doors in terms of future research directions: 
\begin{enumerate}
    \item How effective are these strategies? Are there gaps between how relevant people think they are vs. how effective they actually are? How do we begin to measure actual effectiveness beyond self-reporting?
    \item How do demographic and intersectional factors affect various strategies?
    \item What are other strategies for the barriers observed in this work? How do these strategies complement the strategies observed here?
    \item Who are the stakeholders in these strategies, and what can they do to better support current and future SEs with non-traditional backgrounds? E.g., how can firms better support these SEs and utilize their unique strengths?
\end{enumerate}

\subsection{Advantages due to Non-traditional Backgrounds}
It is clear that technical knowledge is not the only thing required to be successful in SE. Often, less tangible skills like communication skills and maturity that are gained from prior experiences seemed to have played an essential role in obtaining a SE career, beyond programming skills. Interestingly, this aspect of skill transfer was much less obvious in the exploratory study than in the follow-up survey. 
\review{There are multiple likely reasons for this; other than the possible differences between Reddit users whose posts were analyzed and the survey participants (discussed more in Section \ref{sec:thre}), the survey explicitly asked participants about advantages due to having non-traditional backgrounds, while users with non-traditional backgrounds posting to Reddit might be more focused on finding solutions to barriers they faced.}
Moreover, even in threads where people share their successful switching process, this aspect of skill transfer was not discussed as often as the survey participants. This could be because the subreddits studied are still mostly used by SEs with traditional educational and occupational backgrounds, who are less aware of potentially valuable skill transfers from other fields. 
Regardless, recognizing that there are valuable skills that could be advantageous to a SE career might be important when considering whether to switch into SE, which is a decision process that could easily be dominated by considerations related to a person's lack of programming skills. Similarly, helping people harness their existing skills when switching into SE (e.g., by selling these skills better during interviews for SE jobs) could increase the chances of a successful switch. 
On the other hand, companies hiring for SE roles could pay more attention on applicants' soft skills beyond their technical knowledge, especially given the observed trend of underestimations of how important soft skills are by the SE industry \cite{ahmed2012evaluating}.
Moreover, the prominence of skill transfers suggest future research investigating the role of such skills in the formation of mitigation strategies that are personalized to each person's educational and occupational background to be meaningful.

\subsection{Generalizability of Findings and the Theoretical Model}\label{sec:discgeneralize}
\review{
In this subsection, we discuss how transferable findings made in this work are to other contexts beyond the transition process for those with non-traditional educational or occupational backgrounds into a SE career.

\subsubsection{Barriers and Strategies}
\noindent \textbf{Educational} The transition to a SE career has a unique characteristic in that there is a relatively large variety when it comes to educational paths and resources (S1-6).
Other careers might have a higher emphasis on formal education with little to no self-learning methods (e.g., careers related to medical and health), or have less educational resources available online (e.g., careers with very niche skills). 
As such, the combination of high flexibility in learning options, alongside the availability of both formal and informal options make the barriers and strategies less generalizable overall to other fields.

\noindent \textbf{Financial} While the availability of financial aids (S9) is highly personal, as a mitigation strategy, it is probably generalizable to any other contexts where relatively substantial financial costs are involved during the transition process. On that note, we posit that financial costs related to education could serve as a common barrier across fields, but less so for the costs related to job search; paying for access to coding interview preparation platforms is quite specific to the SE context.

\noindent \textbf{Time/Energy} Barriers associated with a lack of time/energy to invest into the transition process, especially factors like work and family strategies (B8-11) are not really unique to SE, and are likely applicable in other career change contexts. On the other hand, the strategy to quit one's current job to focus on learning and job searching (S10) will be more applicable in fields where there might be a significant amount of learning required to change careers.

\noindent \textbf{Psychological} Across all categories, the psychological barriers and strategies identified in this work are probably the most generalizable to other contexts. This is not surprising; researchers found that changing to a different line of work is an event that causes a significant change in life and hence stress, especially for young and middle-aged individuals \cite{sands1980cross}. Moreover, Burks et al. also identified a list of ongoing stressors in everyday life, including many that are relevant to career changing, e.g., decisions about career plans in phase I, decisions about course selection in phase II, and needing work but unable to find a job in phase III \cite{Burks1985}.

\noindent \textbf{Others} We posit that age-related barriers (B14) are somewhat universal when examining career transitions, since changing careers inevitably implies having spent some time in other careers. On the other hand, any uncertainties caused by conflicting opinions (B15) are likely correlated with circumstances where one might be exposed to a huge number of opinions (e.g., careers that are more frequently discussed online).
Moreover, we hypothesize that difficulties finding a job (B17) and networking (B16) are probably more common regardless of field, while difficulties related to keeping up with colleagues after getting a job (B18) are especially relevant within fields where self-learning is possible.
Besides, among all the strategies categorized as others, while the strategy of building a portfolio of personal projects (S17) might only be relevant for more design-related fields (e.g., architecture, design), the other strategies (S14-16,18,19) are likely more generalizable.

\subsubsection{Theoretical Model}
We expect a few aspects of the model proposed in Section \ref{sec:model} to be transferable to other contexts.
These aspects include:
\begin{enumerate}
    \item The general four phase structure of the transition process could be generalizable. However, shifts from phase IV to II might be less common in other contexts. For instance, if there exists careers where the job requirements mainly involve tangible conditions, e.g., a degree, then once someone meets those conditions, they are likely to attain a job without any need to return to the preparation phase.
    \item The general categories of barriers and strategies--educational, financial, time/energy and psychological--might also apply to other contexts. While other categories could emerge for non-SE contexts, we hypothesize that these four categories are common across most contexts, even though the specific barriers and strategies within each category might be different.
    \item The relationship between each type of barrier/strategy and the four phases could also be similar in other contexts. In other words, predicting the severity of barriers and availability of strategies in other contexts could similarly inform the expected difficulty of, and length of time spent in, each phase. For instance, if someone who is well-to-do is transitioning into becoming a doctor from a career completely unrelated to the medical and health field, the preparation to change phase (phase II) might take considerably long due to any required formal education, counterbalanced by the ability to study full-time without the need to work while in school. 
    \item It is also possible that even for non-SE contexts, non-traditional backgrounds can offer useful transferable skills and experiences during career changing. Of course, the exact mix of hard skills, soft skills, experiences and access to job opportunities will vary widely across professions.
\end{enumerate}

}

\section{Threats to Validity}
\label{sec:thre}
\review{In this section, we discuss potential weaknesses of the approach used in this work, what was done to attempt to mitigate these weaknesses, as well as possible alternatives for future works.}
The first threat to validity is the subjectivity involved in the Reddit query building process.
There might be important keywords missing from our Reddit search query, since it is impossible to objectively measure the completeness and quality of the keywords. To address this, we snowballed the group of keywords by starting with a set of keywords that we thought were relevant to our research questions and expanded it as we became more familiar with the dataset.
Moreover, the follow-up survey contributes towards a more comprehensive understanding by supplementing the findings and contributing new perspectives.
\review{That said, future work could employ more advance NLP techniques, for example, using lemmatization when performing keyword-based post filtering, or employing keyword extraction to help with keyword identification.
There are also methodological alternatives to those based on Grounded Theory when analyzing Reddit data, as used in this work's exploratory study. Example alternatives include digital ethnography and computational methods like topic modelling  \cite{proferes2021studying}. 
These methods could have been more useful to generate insights if there exists subreddits that are much more focused on discourse around transitioning into SE, since relevant posts are relatively sparse within existing SE-related subreddits.
Regardless, it would have been beneficial to conduct either a survey or interviews with the Redditors behind identified relevant posts or comments in the dataset; this could allow for a deeper investigation into the various contexts involved in their posts and comments to better understand the significance of various barriers and mitigation strategies. This could be especially useful, since sometimes, comments could be vague or lacking in important information. For example, someone mentioning that they did not face a particular barrier could have been interviewed to understand \textit{why} they did not face it. The same sorts of insights could also have been made for the effectiveness of various mitigation strategies.
In a similar vein, future works could also add follow-up interviews for survey participants to understand the reasons behind their responses. For example, in this work, it would have been insightful for a follow-up interview with survey participants to potentially understand why certain barriers/strategies were rated the way they were, e.g., why did web developers not rate the strategy of ``learning web development as a first step'' as being more relevant.
}

Another threat to our work's external validity is its geographic generalizability. We expect most of the posts and comments analyzed to come from users from the United States (US), since the US accounted for 48.93\% of all traffic to Reddit in the six months leading up to June 2021~\cite{reddittraffic}. 
We also expect participants for our follow-up survey to be mostly located in North America. As such, we do not claim to have findings that are generalizable across all continents but to mostly be valid in the context of North America.
\review{In the exploratory study analysis, background information, such as age and native language, was coded to enhance the context for emerging barriers and mitigation strategies. However, what we observed was that most of the background information was age-related, which resulted in the age-related barriers discussed in Section 4.1.1. Other background details were infrequent, and no additional background-related themes emerged. Future research is encouraged to supplement the findings of this study by exploring other sources of information, which may include more background-related information for further investigations into how specific background-related aspects might influence the significance of barriers during the transition process.
Furthermore, author information was ignored during the coding process, partially due to the existence of posts and comments made by \textit{deleted} users, i.e., users who deleted their account but not their posts. 
As such, even though it was not obvious to the coders that multiple posts had been made by the same individual, it is still plausible, and future work should pay more attention to this, especially in contexts where there is a high likelihood for individuals to make multiple posts, e.g., if there is a culture in a subreddit for Redditors to post update posts every few months.
That said, the existence of deleted users poses a problem, since that hampers efforts to identify identical users across posts. One could choose to remove all posts and comments by deleted users, but that not only reduces the amount of data available to analyze, but also disrupts the discussions since some comments might be removed as well.
This could be an underlying weakness of Reddit-based data.
While Reddit is a rich source of data, future works by larger research teams with more budget can better apply grounded theory's theoretical sampling by preparing to branch into other unexpected data sources to explore other possible aspects of the research questions \cite{Hoda2021}.

Regarding the survey, even though responses were checked for validity (i.e., non-empty, meeting inclusion criteria (Section \ref{sec:participants})), future research should include extra attention checks as well. 
Except for one participant, all survey respondents have some form of post-secondary credential, regardless of its relevance to SE/CS. These participants might not reflect the opinions of those transitioning into SE without any post-secondary education. 
It is important to note that there is a lack of understanding on the average level of educational attainment among those trying to transition into a SE career with a non-traditional educational or occupational background, which is beyond this work's scope.
However, the survey participants in this work could have, on average, higher educational attainment levels than the population's average, leading to biased results. 
For example, prior research found that higher parental human capital could explain a significant portion of an individual's level of educational attainment \cite{belzil2007can}. 
This might bias the survey results by lowering the perceived significance of barriers related to access to human capital (e.g., financial cost of software engineering education).
Regarding the sample size, while the results derived from surveying 46 participants may lack generalizability, they serve as an initial effort to draw attention to this topic. Follow-up studies are recommended to validate our findings.

Furthermore, even though the Reddit data analyzed was between 2017 and 2019, the survey was conducted early 2023. 
The COVID-19 pandemic that occurred between these years led to significant layoffs \cite{techlayoff} and shifts to more remote work \cite{techremote} in the SE industry, which could have affected the survey results by introducing new barriers and worsening existing ones. The survey did not ask participants about the pandemic's impacts on their transition into SE, since that was not this work's goal. No participants mentioned the pandemic explicitly when asked to report other barriers in an open-ended question. However, it is still possible that the relevance of certain barriers were inflated due to the pandemic (e.g., a harder time balancing family responsibilities due to working from home).

}

\section{Conclusion}
\label{sec:conc}
This work analyzed barriers and mitigation strategies of software engineers with non-traditional educational and occupational backgrounds, an important aspect of SE diversity that remains unexplored currently.

The analysis was conducted in two distinct stages. The initial step involved an exploratory study, which entailed analyzing pertinent data from Reddit using a grounded-theory-based approach to identify emerging themes and patterns related to barriers and mitigation strategies. Through this process, we identified more than 30 barriers and mitigation strategies in total across multiple aspects, including educational, financial, time/energy-related, and psychological barriers and strategies. 
Subsequently, a follow-up survey was carried out to supplement the findings from the exploratory study. This involved surveying 46 participants with non-traditional educational and occupational backgrounds who were either in the process of switching to SE or had already made a successful transition.
The survey revealed that most barriers and mitigation strategies were rated as relevant to the switching process, and that experiences in non-SE fields could offer unique skills that are important to performing tasks in SE jobs.

Based on the findings, we propose a theoretical model that breaks down the process of transitioning into a SE career into four phases, and presents which types of barriers and strategies are relevant for each phase, and how non-traditional backgrounds can be beneficial.

\bibliographystyle{ACM-Reference-Format}
\bibliography{main}

\end{document}